\documentclass[journal,draftcls,onecolumn,12pt,twoside]{IEEEtran}
\usepackage[T1]{fontenc}
\usepackage[latin9]{inputenc}
\usepackage{amsthm,mathtools}
\usepackage{amsmath}
\usepackage{bbm} 
\usepackage{graphicx} 
\usepackage{color} 
\usepackage{cite}
\usepackage{algpseudocode}
\usepackage{algorithm}
\usepackage{amssymb}
\usepackage{esint}
\usepackage{algpseudocode}
\usepackage{epstopdf}
\usepackage{multirow}
\usepackage{makecell}
\usepackage{float}
\usepackage{lipsum}
\usepackage{balance}
\usepackage{tikz}
\usepackage{subcaption}

\usetikzlibrary{arrows}
\usetikzlibrary{automata} 
\usetikzlibrary{positioning} 
\tikzstyle{line} = [draw,thick,-latex]
\tikzstyle{transition} = [font=\small]

\theoremstyle{plain}
\theoremstyle{plain}
\newtheorem{definition}{Definition}

\newtheorem{example}{Example}
\newtheorem{theorem}{Theorem}

\newtheorem{result}{Result}

\IEEEoverridecommandlockouts

\definecolor{brown}{rgb}{0.6, 0.2, 0.0}
\definecolor{green}{rgb}{0.13, 0.55, 0.13}

\newif\ifcomment
\commenttrue

\begin{document}

\title{Power Spectra of Constrained Codes with Level-Based Signaling: Overcoming Finite-Length Challenges}

\author{
   \IEEEauthorblockN{Jessica Centers*, Xinyu Tan*, 
   Ahmed Hareedy, \IEEEmembership{Member, IEEE}, \\ and Robert Calderbank, \IEEEmembership{Fellow, IEEE}}
   
   \thanks{*These two authors contributed equally to this work.}
   \thanks{J. Centers, X. Tan, A. Hareedy, and R. Calderbank are with Duke University, Durham, North Carolina, 27708, USA. J. Centers, A. Hareedy, and R. Calderbank are with the Electrical and Computer Engineering Department. X. Tan is with the Computer Science Department. (E-mail: jessica.bilskie@duke.edu, xinyu.tan@duke.edu, ahmed.hareedy@duke.edu, and robert.calderbank@duke.edu).}\vspace{-2.0em}
}
\maketitle

%%%%%%%%%%%%%%%%%%%%%%%%%%%%%%%%%%%%%
\begin{abstract}
In various practical systems, certain data patterns are prone to errors if written or transmitted. In magnetic recording and communication over transmission lines, data patterns causing consecutive transitions that are not sufficiently separated are prone to errors. In Flash memory with two levels per cell, data patterns causing high--low--high charge levels on adjacent cells are prone to errors. Constrained codes are used to eliminate error-prone patterns, and they can also achieve other goals. Recently, we introduced efficient binary symmetric lexicographically-ordered constrained (LOCO) codes and asymmetric LOCO (A-LOCO) codes to increase density in magnetic recording systems and lifetime in Flash systems by eliminating the relevant detrimental patterns. Due to their application, LOCO and A-LOCO codes are associated with level-based signaling. Studying the power spectrum of a random signal with certain properties is principal for any storage or transmission system. It reveals important properties such as the average signal power at DC, the bandwidth of the signal, and whether there are discrete power components at certain frequencies. In this paper, we first modify a framework from the literature in order to introduce a method to derive the power spectrum of a sequence of constrained data associated with level-based signaling. We apply our method to infinitely long sequences satisfying symmetric and asymmetric constraints. Next, we show how to generalize the method such that it works for a stream of finite-length codewords as well, thus demonstrating how to overcome the associated finite-length challenges. We use the generalized method to devise closed forms for the spectra of finite-length LOCO and A-LOCO codes from their transition diagrams. Our LOCO and A-LOCO spectral derivations can be performed for any code length and can be extended to other constrained codes. We plot these power spectra, and discuss various important spectral properties for both LOCO and A-LOCO codes. We also briefly discuss an alternative method for deriving the power spectrum, and introduce an idea towards reaching the spectra of self-clocked codes.
\end{abstract}

\begin{IEEEkeywords}
Power spectra, constrained codes, finite-length, LOCO codes, data storage, data transmission.
\end{IEEEkeywords}

%\tableofcontents

%%%%%%%%%%%%%%%%%%%%%%%%%%%%%%%%%%%%%
\section{Introduction}\label{sec_intro}

Shannon introduced the concept of coding for constrained channels in 1948 \cite{shan_const}. He represented a constrained sequence via a finite-state transition diagram (FSTD) and showed how the capacity can be derived. About $20$ years later, researchers started to use constrained codes in magnetic recording (MR) devices adopting peak detection to increase their density \cite{tang_bahl, franaszek}. These codes were $(d, k)$ binary run-length-limited (RLL) codes, which control the minimum (resp., maximum) separation between consecutive transitions via the parameter $d$ (resp., $k$) to mitigate inter-symbol interference (ISI) (resp., enhance self-clocking) \cite{siegel_mr, immink_surv}. RLL codes are associated with transition-based signaling, which is bipolar \textit{non-return-to-zero inverted (NRZI) signaling} in the binary case. In NRZI signaling, a $0$ is represented by no transition, while a $1$ is represented by a transition. Constrained codes are still being used in modern MR systems adopting sequence detection \cite{vasic_prc, ahh_scmr}, which is demonstrated in \cite{siegel_const} and \cite{ahh_loco}.

Constrained codes find application in Flash memory devices to help increase their lifetime. A Flash cell is a MOSFET transistor with a floating gate, the amount of charge in which controls its operation. Consider three adjacent Flash cells, where the outer two are to be programmed to the highest charge level, while the inner one is to be programmed to a lower level. Charges propagate from the outer cells to the inner one, resulting in inter-cell interference (ICI) that can corrupt the stored data \cite{lee_ici}. For Flash devices with $2$ levels per cell, codes preventing the pattern $101$ were presented in \cite{qin_flash} and \cite{ahh_aloco}. For Flash devices with more than $2$ levels per cell, efficient constrained codes were introduced in recent literature \cite{veeresh_mlc, ahh_qaloco}. These codes are associated with level-based signaling, which is bipolar \textit{non-return-to-zero (NRZ) signaling} in the binary case. In NRZ signaling, a $0$ is represented by a level, while a $1$ is represented by another level. In this paper, the two levels are always $-1$ for $0$ and $+1$ for $1$, respectively.

There are various other applications of constrained codes. Two-dimensional magnetic recording (TDMR) is a new technology that promises a boast in the density of MR devices \cite{wood_tdmr, chan_tdmr}. Constrained codes preventing certain, error-prone TD patterns increase the reliability of TDMR devices \cite{bd_tdmr}. Constrained codes find application in optical recording devices \cite{immink_opt}. Constrained codes are also used in various data transmission systems, as they can mitigate cross-talk between wires over which the data is transmitted \cite{sridhara_ctalk}, and they can achieve DC-balance, i.e., zero average power at frequency zero \cite{saade_comp}.

Even though RLL codes presented in \cite{tang_bahl} were based on lexicographic indexing, the coding community diverted to codes based on finite-state machines (FSMs) shortly after that \cite{siegel_mr, ach_fsm}. However, it is a complicated task to design FSM-based constrained codes that are capacity-approaching, and the task becomes even more complicated for high rate codes \cite{ahh_qaloco}. Constrained codes based on lexicographic indexing were presented in \cite{immink_lex}, which is guided by \cite{cover_lex}. Recently, we returned to the fundamentals of constrained coding \cite{shan_const, tang_bahl}, and designed binary symmetric lexicographically-ordered constrained (LOCO) codes primarily for magnetic recording applications to help increase density \cite{ahh_loco}. We then designed binary asymmetric LOCO (A-LOCO) codes for single-level cell (SLC) Flash devices, i.e., with $2$ levels per cell, to help increase lifetime \cite{ahh_aloco}. We also developed constrained codes for Flash devices with more than $2$ levels per cell \cite{ahh_qaloco} and for TDMR devices \cite{bd_tdmr}. All our codes are capacity-achieving, offer simple encoding-decoding, and reconfiguring them is as simple as reprogramming an adder.

Given the set of patterns to forbid, one can design an infinite sequence as well as a finite-length block code satisfying the constraint. In data storage and data transmission, deriving the power spectral density (PSD) of a random stream of data/codewords constrained by forbidding error-prone patterns after signaling is quite important in the infinite and more importantly the finite-length scenarios. The PSD reveals to the system designer how power is distributed across different frequencies \cite{immink_book, gall_psd}. Consequently, the system designer identifies pivotal properties such as the average power of the stream at DC, the stream bandwidth, i.e., the frequency range in which most of the power is allocated, in addition to the discrete power components, if any, which result from the periodicity of the auto-correlation function \cite{immink_book}.

There is a rich literature on studying power spectra of signals in various systems. In the context of data transmission, power spectra of coded streams generated by an FSM and transmitted over digital repeatered lines were discussed in \cite{bosik_psd} and \cite{cairo_psd}. In the context of wireless communications, numerical \cite{justes_psd} and anayltical \cite{piment_psd} methods for obtaining the spectra of coded streams generated by an FSM were presented. The power spectrum of a block-coded modulated signal having a multi-dimensional constellation was discussed in \cite{biglier_psd}. In the context of data storage, power spectra of certain FSM-based codes in MR systems were derived in \cite{lind_psd}, and spectral null analysis of $(d, k)$ RLL codes in MR and magneto-optic systems was presented in \cite{meny_psd}. The effect of asymmetry in coded data associated with NRZ signaling on the power spectrum was detailed in \cite{bishop_psd}. Finally, \cite{gall_psd}, which we shortly refer to again in this paper, introduced a method to find power spectra of RLL codes associated with NRZI signaling for storage and transmission, emphasizing more on infinite sequences.

In this paper, we start from the method in \cite{gall_psd} and devise a method for obtaining the power spectrum of a binary constrained stream that is associated with NRZ signaling. Let $\bold{y}^r$ be a run of $r$ consecutive $y$'s. Define $\mathcal{S}_x = \{010, \allowbreak 101, 0110, 1001, \dots, 0\bold{1}^x0, 1\bold{0}^x1\}$, which is symmetric, i.e., closed under taking pattern complements, and define $\mathcal{A}_x = \{101, 1001, \dots, 1\bold{0}^x1\}$, which is asymmetric. We use our method to obtain the power spectra of infinitely long $\mathcal{S}_x$-constrained and $\mathcal{A}_x$-constrained sequences. We then generalize this method for streams of finite-length constrained codewords, illustrating how to work with transition diagrams that depend on the code length and how to move from wide-sense-stationary into cyclostationary random processes. We apply the new generalized model to obtain the spectra of LOCO and A-LOCO codes, which are binary finite-length constrained codes. Our new generalized method can also be used for other constrained codes, and its complexity growth with the code length is limited. We use the plots of power spectra for various code parameters to understand pivotal properties about LOCO and A-LOCO coded streams such as the power at DC and the $3$dB bandwidth. Finally, we verify our spectral derivations for $\mathcal{S}_x$-constrained and $\mathcal{A}_x$-constrained sequences via a different method, and take a step towards obtaining the spectra of self-clocked LOCO and A-LOCO codes.

The rest of the paper is organized as follows. In Section~\ref{sec_generalNRZ}, we introduce our method for deriving the spectrum of a constrained code with NRZ signaling. In Section~\ref{sec_infinite}, we find closed-form expressions for the PSDs of infinitely long constrained codes. In Section~\ref{sec_finiteFSTD}, we show how to generate transition diagrams for finite-length codes. In Section~\ref{sec_finite}, we demonstrate how to deal with cyclostationary processes, and we find closed-form expressions for the PSDs of A-LOCO and LOCO codes. We discuss important properties of these power spectra. In Section~\ref{sec_conc}, we conclude the paper. The appendices discuss an alternative method and self-clocked codes.

%%%%%%%%%%%%%%%%%%%%%%%%%%%%%%%%%%%%%
\section{Deriving the PSD for a Constrained Code With NRZ Signaling}\label{sec_generalNRZ}

Gallopoulos et al. presented a method to calculate the spectrum of an RLL code with NRZI signaling based on its finite-state transition diagram (FSTD)~\cite{gall_psd}. We modify their method to address lexicographically-ordered (see Section~\ref{sec_finiteFSTD}) as well as FSM-based constrained codes associated with NRZ signaling. We note that the FSTDs we discuss in this paper represent each state as a fixed number of the most recently generated bits.

We first derive the one-step state transition matrix (OSTM) via the following steps:
\begin{enumerate}
    \item Derive the FSTD that generates a valid code sequence $\{X_n\}$ of binary data satisfying the system constraints and keep track of the probabilities of each transition. Make sure that each edge generates a single symbol (either a $1$ or $0$) and that all the incoming edges for a given vertex are generated by the same symbol.
    \item Derive the one-step state transition diagram (OSTD) of the same system by assigning states only to the vertices where the most recent generated bit is a $1$. The edges of the OSTD define the number of transitions (or symbols) it takes to get from one state to the next according to the FSTD. It is necessary to maintain record of the probabilities associated with these edges. We define the sequence produced by the OSTD as $\{T_i\}$ (iid). In particular, $T_i$ represents the number of consecutive $0$'s along with the ending $1$ of the $i$th run in $\{X_n\}$.
    \item Generate the one-step state transition matrix (OSTM) ${\bf{G}}(D)$ from the known edge probabilities and run lengths defining the OSTD. The general entry ${\bf{G}}_{ij}(D)$ is given by
    \begin{align}\label{eq:def_G_D}
        {\bf{G}}_{ij}(D) = \sum_{t=1}^\infty p_{ij}(t)D^t,
    \end{align}
    where $p_{ij}(t)$ is the probability of transiting from state $i$ to state $j$ in the OSTD and $t$ is the run length characterizing that transition.
\end{enumerate}
We then derive the closed-form expression for the power spectral density (PSD) of a written or transmitted sequence using the OSTM. We adopt the following signal generation scheme:
\begin{align}
    \text{Code, } \{X_n\}\xrightarrow[\text{signaling}]{\text{NRZ}}\text{Modulation sequence,  }\{Y_n\}\xrightarrow[\text{shaping}]{\text{Pulse}} \text{Write signal,  }W(t).
\end{align}
The PSD function of a wide-sense-stationary discrete-time process $\{X_n\}$ is given by
\begin{align}
    S_X(D) = \sum_{k=-\infty}^\infty \mathbb{R}_X(k)D^k = \sum_{k=-\infty}^\infty \mathbb{E}[X_0X_k]D^k,
\end{align}
where $\mathbb{R}_X(k)$ is the auto-correlation function with time lag $k$, and it is an even function. $D$ is defined as the complex exponential $e^{i2\pi fT}$ for some frequency $f$ and bit interval $T$. In the case that the process $\{X_n\}$ is cyclostationary (aside from the cyclic properties in $W(t)$ that arise from the pulse shaping), the PSD is found by obtaining its continuous and discrete portions separately. We will provide discussion and details on this topic in Section~\ref{sec_finite}.

We first present a fundamental result that is analogous to \cite[Theorem~1]{gall_psd}. Vectors are by default row vectors unless otherwise stated.

\begin{theorem}\label{thm:1}
Assume that the Markov chain describing the code sequence generation is at equilibrium. The power spectrum $S_X(D)$ of the process $\{X_n\}$ is given by
\begin{align}\label{eq:thm_Sx(D)}
    S_X(D) = p(1)\boldsymbol{\pi}\left[({\bf{I}} - {\bf{G}}(D))^{-1} + ({\bf{I}} - {\bf{G}}(D^{-1}))^{-1} - {\bf{I}}\right]{{\bf u}}^\mathsf{T},
\end{align}
where ${\bf{u}}$ is an all-one vector, $p(1)$ is the equilibrium probability of a $1$ in $\{X_n\}$, ${\bf{I}}$ is the identity matrix of the same size as ${\bf{G}}(D)$, and $\boldsymbol{\pi}$ is the stationary distribution.
\end{theorem}
\begin{IEEEproof}
The stationary distribution $\boldsymbol{\pi}$ satisfies $\boldsymbol{\pi} {\bf u}^\mathsf{T}=1$ and $\boldsymbol{\pi} {\bf{G}}(1)=\boldsymbol{\pi}$, where ${\bf{G}}(1)$ is the ordinary OSTM of the Markov chain. 
We connect ${\bf{G}}(D)$ with $S_X(D)$. Define
\begin{align}\label{eq:def_Phi}
    \Phi_l(D) = \sum_{k=0}^\infty D^k \cdot \mathbb{P}[\text{length of }l\text{ consecutive }T_i\text{ is }k|\text{run started}].
\end{align}
It follows from (\ref{eq:def_G_D}) and the definition of $\{T_i\}$ that (see also \cite{gall_psd})
\begin{align}\label{eq:Phi_D_and_G_D}
    \Phi_l(D) = \boldsymbol{\pi} [{\bf{G}}(D)]^l {\bf u}^\mathsf{T} \text{ and }\mathbb{E}[T_i]=\frac{1}{p(1)}=\boldsymbol{\pi} {\bf{G}}'(1){\bf u}^\mathsf{T},
\end{align}
where ${\bf{G}}'(1)$ is the derivative of ${\bf{G}}(D)$ with respect to $D$ when $D=1$. Then,
\begin{align}
    \sum_{l=0}^\infty \Phi_l(D) &= \sum_{k=0}^\infty D^k \left( \sum_{l=0}^\infty \mathbb{P}[\text{length of }l\text{ consecutive }T_i\text{ is }k|\text{run started}] \right) \nonumber\\
    &= 1 + \sum_{k=1}^\infty D^k \cdot \mathbb{P}[X_k = X_0 = 1|X_0 = 1] \nonumber\\
    &= 1+\sum_{k=1}^\infty D^k \cdot \mathbb{E}[X_0 X_k|X_0 = 1].
\end{align}
Therefore, the power spectral density $S_X(D)$ of $\{X_n\}$ is
\begin{align}
    S_X(D) &= \sum_{k = -\infty}^{\infty} D^k \cdot \mathbb{E}[X_0X_k] = p(1)\left(\sum_{l=0}^\infty \Phi_l(D) +\sum_{l=0}^\infty \Phi_l(D^{-1}) - 1\right).
\end{align}
It follows from (\ref{eq:def_Phi}) that $\Phi_l(D) = [\Phi_1(D)]^l = [\Phi(D)]^l $. Thus, using the power series
$$\sum_{l = 0}^\infty x^l = \frac{1}{1-x}, \textup{ for } \vert x \vert < 1,$$
along with (\ref{eq:Phi_D_and_G_D}), we get
\begin{align}\label{eq:Sx(D)}
    S_X(D) &=p(1)\left[\frac{1}{1-\Phi(D)} + \frac{1}{1-\Phi(D^{-1})} - 1\right] \nonumber\\
    & = p(1)\boldsymbol{\pi}\left[({\bf{I}} - {\bf{G}}(D))^{-1} + ({\bf{I}} - {\bf{G}}(D^{-1}))^{-1} - {\bf I}\right]{\bf u}^\mathsf{T},
\end{align}
which completes the proof.
\end{IEEEproof}

After NRZ signaling, we have the modulation sequence $\{Y_n\}$ with $Y_n = 2X_n - 1$. Thus, for any $k\in\mathbb{Z}$,
\begin{align}\label{eq:exp_y}
    \mathbb{R}_Y(k) = \mathbb{E}[Y_0 Y_k] = \mathbb{E}[(2X_0-1) (2X_k-1)] =4\mathbb{E}[X_0X_k] - 4p(1) + 1.
\end{align}
Therefore, the power spectral density $S_Y(D)$ of $\{Y_n\}$ is
\begin{align}
    S_Y(D) &= \sum_{k=-\infty}^{\infty} D^k \cdot \mathbb{E}[Y_0Y_k] \nonumber\\
    & = 4S_X(D) + \left[1-4p(1)\right]\sum_{k = -\infty}^{\infty} D^k. \nonumber
\end{align}
As a result, we get
\begin{align}\label{eq:Sy(D)}
S_Y(D) = \begin{cases}
    4S_X(D), & D \neq 1,\\
    4S_X(D) + 1 - 4p(1), & D = 1.
    \end{cases}
\end{align}

The write signal $W(t)$ is given by
\begin{align}
    W(t) = \sum_{n=-\infty}^\infty Y_n P_T(t-nT),
\end{align}
where $P_T(t)$ is the modulation pulse function (chosen as a rectangular pulse in the context of this work) with bit interval $T$, i.e.,
\begin{align}
    P_T(t)=\begin{cases}
    1, & 0\leq t<T,\\
    0, & \text{otherwise}.
    \end{cases}
\end{align}
Therefore, the power spectral density $S_W(f)$ of the write signal $W(t)$ is given by
\begin{align}\label{eq:Sw(f)}
    S_W(f)=\text{sinc}^2(\pi fT) T^2 S_Y(e^{i2\pi fT}).
\end{align}
For $f=0$, which is equivalent to $D =1$, (\ref{eq:Sw(f)}) becomes
\begin{equation}\label{eq:Sw(0)}
S_W(0)=T^2 S_Y(1)=T^2\left(4S_X(1) + 1 - 4p(1)\right).
\end{equation}
In the following, we consider $f$ to be the normalized frequency, i.e., $T$ is set to $1$.
% \begin{enumerate}
%     \item Calculate the vector of stationary probabilities, $\pi$, from the following two equations:
%         \begin{itemize}
%             \item $\pi G(1)=\pi$ : where G(1) is the ordinary
%         one-step state transition matrix of the Markov chain  
%             \item $\sum_{i=1}^{L}\pi_{i}=1$ : where L is the total number of states in the run-length diagram derived
%         \end{itemize}
%     \item  Calculate the equilibrium probability of the Markov Chain, $p(1)$, where $p(1)=[\pi G'(1)u^{T}]^{-1}$ and $G'(D)$ is the derivative of $G(D)$ with respect to $D$.
%     \item Calculate the spectrum of an intermediate signal. Theorem 1 in \cite{gall_psd} demonstrates that the spectrum of their intermediate signal $X(t)$ for NRZI signaling is found by:
%         \begin{itemize}
%             \item $S_{X}(D)=p(1)\pi [(I+G(D))^{-1}+(I+G(D^{-1}))^{-1}-I]u^{T}$
%         \end{itemize}
%     \item With $S_{X}$, the spectrum of the NRZ write signal $S_{W}(f)$ can then be found via the following two equations:
%         \begin{itemize}
%             \item $S_{X}(D)=\frac{2-(D+D^{-1})}{4}S_{A}(D)$ : for $D=e^{i2\pi f}$, then $S_{X}(f)=(sin(\pi f))^{2}S_{A}(f)$
%             \item $S_{W}(f)=(\frac{sin(\pi fT)}{\pi fT})^{2}T^{2}S_{A}(f)$ but for $f=0$, $S_{W}(f)=T^{2}S_{A}(f)$
%         \end{itemize}
% \end{enumerate}

In this paper, for any coding schemes except LOCO codes, we obtain the PSD through~(\ref{eq:thm_Sx(D)}),~(\ref{eq:Sy(D)}), (\ref{eq:Sw(f)}), and~(\ref{eq:Sw(0)}). Given that LOCO codes require a three-level waveform because of their bridging \cite{ahh_loco}, we modify the general procedure in Subsection~\ref{subsec_LOCO}.

% %%%%%%%%%%%%%%%%%%%%%%%%%%%%%%%%%%%%%
\section{Closed-Form PSD of Infinite-Length Constrained Codes}\label{sec_infinite}

Consider the following two infinite-length codes constrained by the separation of bit transitions: $\mathcal{A}_x$-constrained codes and $\mathcal{S}_x$-constrained codes. Though technically the transition separation is constrained only when the first transition occurs from a $1$ to a $0$ for $\mathcal{A}_x$-constrained codes, we still categorize them into the family of codes constrained by transition separation. Code sequences here represent wide-sense stationary processes.

\subsection{PSD of Infinite-Length $\mathcal{A}_x$-Constrained Codes}\label{subsec_Ax}

\begin{definition}\label{def:Ax_set}
Let $\mathcal{A}_x=\{101, 1001, \dots, 1{\bf 0}^x1\}$ with $|\mathcal{A}_x| = x$. Define a binary $\mathcal{A}_x$-constrained code which forbids any pattern in $\mathcal{A}_x$ from appearing in the coded sequences. 
\end{definition}
$\mathcal{A}_{X}$-constrained codes are used in Flash memory systems to eradicate data patterns, such as $101$, that cause parasitic capacitances to result in maximum charge propagation across cells during the programming phase (ICI). These codes are typically associated with NRZ signaling. Thus, Theorem~\ref{thm:1} is applicable when calculating the PSD for this class of codes.

\begin{example}\label{eg:A_1}
We first provide an example of deriving the OSTM for infinite-length $\mathcal{A}_1$-constrained code, which forbids $101$ in the code sequence. Using mathematical induction, the interested reader can finish the derivation of the OSTM for any $\mathcal{A}_x$-constrained code (see Result~\ref{res:OSTM_A}). 

Fig.~\ref{subfig:OSTM_A1_a} shows the FSTD of $\mathcal{A}_1$-constrained code, where each edge is labeled by its single binary symbol transition. Fig.~\ref{subfig:OSTM_A1_b} records the probability for each corresponding transition, and it also labels in gray only the states resulting from a $1$ transition. 

\tikzset{node distance=1.4cm,
every state/.style={semithick,minimum size=1pt},
initial text={},
double distance=2pt,
every edge/.style={ draw,
->,>=stealth', auto, semithick}}

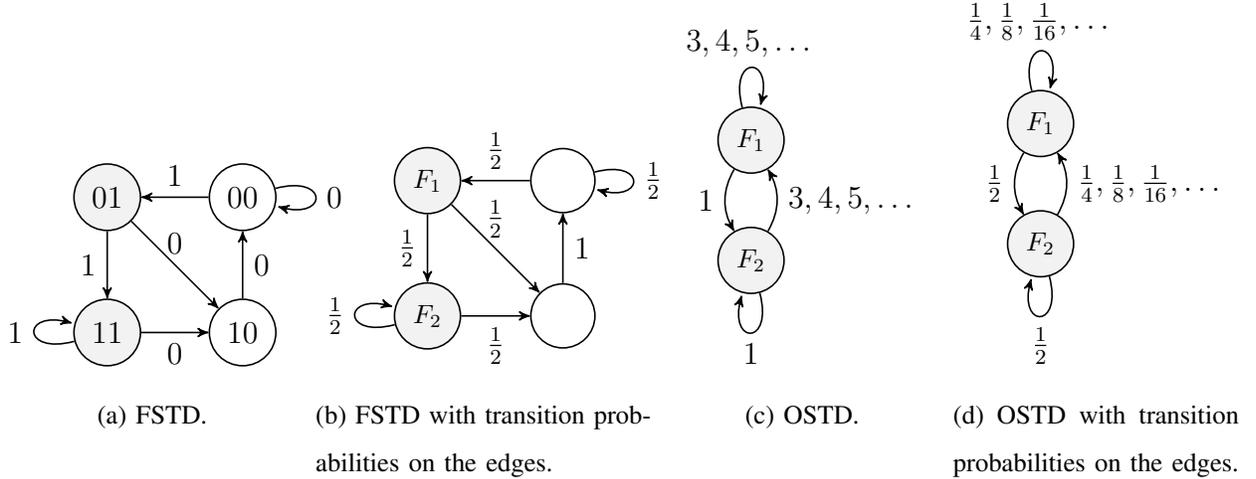
\begin{figure}
\begin{subfigure}[t]{.25\textwidth}
\centering
\begin{tikzpicture}[node distance=1.8cm]
\node[state] (A) {$00$};
\node[state, fill=gray!10, left of=A] (B) {$01$}; 
\node[state, fill=gray!10, below of=B] (C) {$11$};
\node[state, right of=C] (D) {$10$};
\draw (A) edge[loop right] node {$0$} (A);
\draw (A) edge node[above] {$1$} (B);
\draw (B) edge node[left] {$1$} (C);
\draw (B) edge node[above] {$0$} (D);
\draw (C) edge[loop left] node {$1$} (C);
\draw (C) edge node[below] {$0$} (D);
% \draw (D) edge[bend right=30] node[right] {0} (A);
\draw (D) edge node[right] {$0$} (A);
\end{tikzpicture}
\caption{FSTD.}\label{subfig:OSTM_A1_a}
\end{subfigure}
\begin{subfigure}[t]{.27\textwidth}
\centering
\begin{tikzpicture}[node distance=1.8cm]
\node[state] (A) {\textcolor{white}{$D$}};
\node[state, fill=gray!10, left of=A] (B) {\small$F_1$}; 
\node[state, fill=gray!10, below of=B] (C) {\small$F_2$};
\node[state, right of=C] (D) {\textcolor{white}{$D$}};
\draw (A) edge[loop right] node {$\frac{1}{2}$} (A);
\draw (A) edge node[above] {$\frac{1}{2}$} (B);
\draw (B) edge node[left] {$\frac{1}{2}$} (C);
\draw (B) edge node[above] {$\frac{1}{2}$} (D);
\draw (C) edge[loop left] node {$\frac{1}{2}$} (C);
\draw (C) edge node[below] {$\frac{1}{2}$} (D);
\draw (D) edge node[right] {$1$} (A);
\end{tikzpicture}
\caption{FSTD with transition probabilities on the edges.}\label{subfig:OSTM_A1_b}
\end{subfigure}
\begin{subfigure}[t]{.23\textwidth}
\centering
\begin{tikzpicture}[node distance=1.6cm]
\node[state, fill=gray!10] (B) {\small$F_1$}; 
\node[state, fill=gray!10, below of=B] (C) {\small$F_2$};
\draw (B) edge[bend right] node[left] {$1$} (C);
\draw (B) edge[loop above] node[above] {$3,4,5,\dots$} (B);
\draw (C) edge[loop below] node[below] {$1$} (C);
\draw (C) edge[bend right] node[right] {$3,4,5,\dots$} (B);
\end{tikzpicture}
\caption{OSTD.}\label{subfig:OSTM_A1_c}
\end{subfigure}
\begin{subfigure}[t]{.23\textwidth}
\centering
\begin{tikzpicture}[node distance=1.6cm]
\node[state, fill=gray!10] (B) {\small$F_1$}; 
\node[state, fill=gray!10, below of=B] (C) {\small$F_2$};
\draw (B) edge[bend right] node[left] {$\frac{1}{2}$} (C);
\draw (B) edge[loop above] node[above] {$\frac{1}{4}, \frac{1}{8}, \frac{1}{16}, \dots$} (B);
\draw (C) edge[loop below] node[below] {$\frac{1}{2}$} (C);
\draw (C) edge[bend right] node[right] {$\frac{1}{4}, \frac{1}{8}, \frac{1}{16}, \dots$} (B);
\end{tikzpicture}
\caption{OSTD with transition probabilities on the edges.}\label{subfig:OSTM_A1_d}
\end{subfigure}
\caption{FSTD and OSTD for infinite-length $\mathcal{A}_1$-constrained sequence.} \label{fig:OSTM_A1}
\end{figure}

We obtain Fig.~\ref{subfig:OSTM_A1_c} by using only the labeled states and drawing all the possible transitions from one state to another. The edges are labeled by run-length(s) $t_i$, indicating a transition corresponding to a sequence ${\bf 0}^{t_i-1}1$. We also keep track of the probabilities associated with the edge labels in the OSTD in Fig.~\ref{subfig:OSTM_A1_d}. It then follows from~(\ref{eq:def_G_D}) that the OSTM of infinite-length $\mathcal{A}_1$-constrained code is given by
\begin{align}
    {\bf{G}}(D) = \begin{bmatrix}
    \alpha & \frac{1}{2}D\\
    \alpha & \frac{1}{2}D
    \end{bmatrix},
\end{align}
where
\begin{align}
    \alpha = \sum_{k=0}^\infty \left(\frac{1}{2}\right)^{2+k}D^{3+k} = \frac{D^3}{2(2-D)}.
\end{align}
\end{example}

\begin{result}
\label{res:OSTM_A}
According to our derivation by induction, the OSTM for any $\mathcal{A}_x$-constrained code is then
\begin{align}
    {\bf{G}}(D)=\begin{bmatrix} 
        \begin{matrix}
        \alpha \\ \vdots
        \end{matrix} & \frac{1}{2}D{\bf{I}}_x \\
        \alpha & \begin{matrix}
        0 & \cdots & 0 & \frac{1}{2}D
        \end{matrix}
        \end{bmatrix},
\end{align}
%  \begin{pspicture}
%  $ \begin{pmatrix}
%     \pnode[-0.3ex, 1.5ex]{A}
%     a_{11} & a_{12} & a_{13} & b_{11} & b_{12} \\
%     a_{21} & a_{22} & a_{23} & b_{21} & b_{22} \\
%     c_{11} & c_{12} &
%     c_{13} \pnode[1.4ex, -0.8ex]{C} & d_{11} & d_{12} \\
%     c_{21} & c_{22} & c_{23} & d_{21} & d_{22} \\
%     c_{31} & c_{32} & c_{33} & d_{31} & d_{32}\pnode[0.2ex, -0.6ex]{D}\,
%   \end{pmatrix} $
% \psset{linecolor = red}
% \psframe(A)(C)
% \psset{dimen = inner}\psframe(C)(D)
% \end{pspicture}
where 
\begin{align}
    \alpha = \sum_{k=0}^{\infty}\left(\frac{1}{2}\right)^{2+k}D^{2+x+k}=\frac{D^{2+x}}{2(2-D)}
\end{align}
and $\bf{I}_x$ is the identity matrix of size $x\times x$.
\end{result}

\begin{figure} 
\vspace{-0.9em}
\centering
\includegraphics[width=10cm]{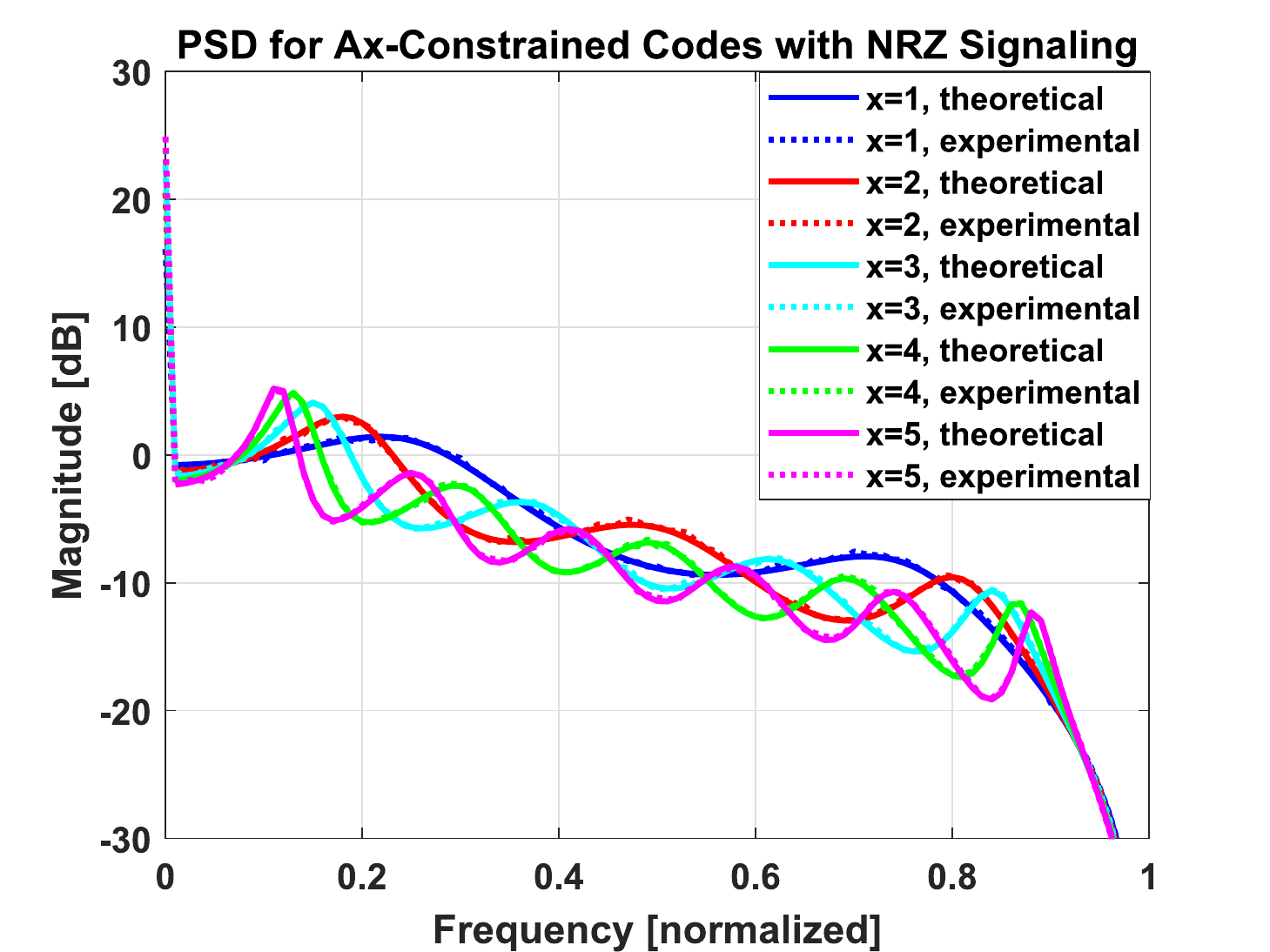}
\caption{PSDs for infinite-length $\mathcal{A}_x$-constrained codes with NRZ signaling. The spike at DC indicates the expected discrete DC power component due to asymmetry.}\label{fig:Ax_PSD}
\vspace{-0.7em}
\end{figure}

We then generate the PSD for an infinite-length $\mathcal{A}_x$-constrained code following~(\ref{eq:thm_Sx(D)}),~(\ref{eq:Sy(D)}),~(\ref{eq:Sw(f)}), and~(\ref{eq:Sw(0)}). Fig.~\ref{fig:Ax_PSD} shows the PSDs for these codes with $x \in \{1, 2, \dots, 5\}$.

\subsection{PSD of Infinite-Length $\mathcal{S}_x$-Constrained Codes}\label{subsec_Sx}

\begin{definition}\label{def:Sx_set}
Let $\mathcal{S}_x=\{010, 101, 0110, 1001, \dots, 0{\bf 1}^x0, 1{\bf 0}^x1\}$ with $|\mathcal{S}_x| = 2x$. Define a binary $\mathcal{S}_x$-constrained code which forbids any pattern in $\mathcal{S}_x$ from appearing in the coded sequences. 
\end{definition}

$\mathcal{S}_x$-constrained codes forbid data patterns that contribute the most to ISI in magnetic recording systems, i.e., the most error-prone data patterns in these systems. They find another application in some standards for data transmission. These codes are also typically associated with NRZ signaling, thus justifying the use of Theorem~\ref{thm:1} to calculate their PSDs.

\begin{example}
In a way similar to Example~\ref{eg:A_1}, we will derive the OSTM for infinite-length $\mathcal{S}_1$-constrained code, which forbids $010$ and $101$ in the code sequence. The interested reader can again finish the derivation of the general OSTM using mathematical induction (see Result~\ref{res:OSTM_S}).

Fig.~\ref{subfig:OSTM_S1_a} shows the FSTD of $\mathcal{S}_1$-constrained code where each edge is labeled by its single binary symbol transition. Fig.~\ref{subfig:OSTM_S1_b} records the probability for each corresponding transition, and it also labels in gray only the states resulting from a $1$ transition. 

\tikzset{node distance=1.4cm,
every state/.style={semithick,minimum size=1pt},
initial text={},
double distance=2pt,
every edge/.style={ draw,
->,>=stealth', auto, semithick}}

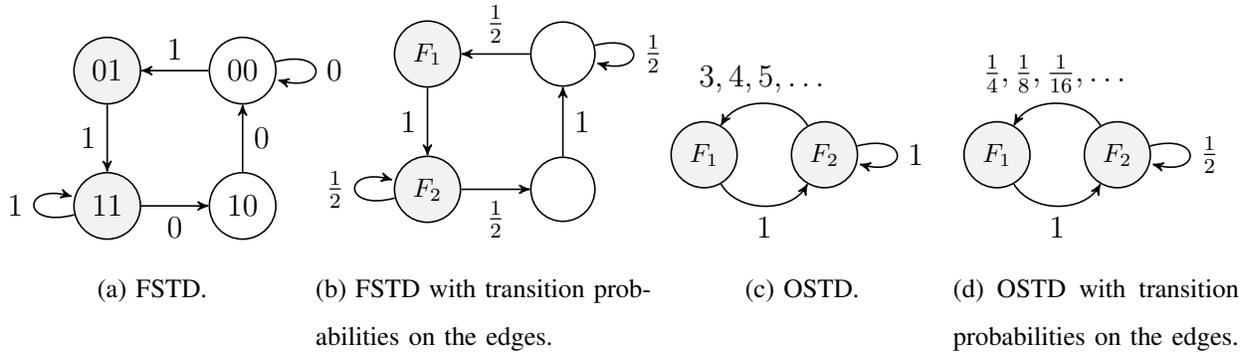
\begin{figure}
\begin{subfigure}[t]{.25\textwidth}
\centering
\begin{tikzpicture}[node distance=1.8cm]
\node[state] (A) {$00$};
\node[state, fill=gray!10, left of=A] (B) {$01$}; 
\node[state, fill=gray!10, below of=B] (C) {$11$};
\node[state, right of=C] (D) {$10$};
\draw (A) edge[loop right] node {$0$} (A);
\draw (A) edge node[above] {$1$} (B);
\draw (B) edge node[left] {$1$} (C);
\draw (C) edge[loop left] node {$1$} (C);
\draw (C) edge node[below] {$0$} (D);
% \draw (D) edge[bend right=30] node[right] {0} (A);
\draw (D) edge node[right] {$0$} (A);
\end{tikzpicture}
\caption{FSTD.}\label{subfig:OSTM_S1_a}
\end{subfigure}
\begin{subfigure}[t]{.27\textwidth}
\centering
\begin{tikzpicture}[node distance=1.8cm]
\node[state] (A) {\textcolor{white}{$D$}};
\node[state, fill=gray!10, left of=A] (B) {\small$F_1$}; 
\node[state, fill=gray!10, below of=B] (C) {\small$F_2$};
\node[state, right of=C] (D) {\textcolor{white}{$D$}};
\draw (A) edge[loop right] node {$\frac{1}{2}$} (A);
\draw (A) edge node[above] {$\frac{1}{2}$} (B);
\draw (B) edge node[left] {$1$} (C);
\draw (C) edge[loop left] node {$\frac{1}{2}$} (C);
\draw (C) edge node[below] {$\frac{1}{2}$} (D);
\draw (D) edge node[right] {$1$} (A);
\end{tikzpicture}
\caption{FSTD with transition probabilities on the edges.}\label{subfig:OSTM_S1_b}
\end{subfigure}
\begin{subfigure}[t]{.23\textwidth}
\centering
\begin{tikzpicture}[node distance=1.6cm]
\node[state, fill=gray!10] (B) {\small$F_1$}; 
\node[state, fill=gray!10, right of=B] (C) {\small$F_2$};
\draw (B) edge[bend right=60] node[below] {$1$} (C);
\draw (C) edge[loop right] node {$1$} (C);
\draw (C) edge[bend right=60] node[above] {$3,4,5,\dots$} (B);
\end{tikzpicture}
\caption{OSTD.}\label{subfig:OSTM_S1_c}
\end{subfigure}
\begin{subfigure}[t]{.23\textwidth}
\centering
\begin{tikzpicture}[node distance=1.6cm]
\node[state, fill=gray!10] (B) {\small$F_1$}; 
\node[state, fill=gray!10, right of=B] (C) {\small$F_2$};
\draw (B) edge[bend right=60] node[below] {$1$} (C);
\draw (C) edge[loop right] node {$\frac{1}{2}$} (C);
\draw (C) edge[bend right=60] node[above] {$\frac{1}{4}, \frac{1}{8}, \frac{1}{16}, \dots$} (B);
\end{tikzpicture}
\caption{OSTD with transition probabilities on the edges.}\label{subfig:OSTM_S1_d}
\end{subfigure}
\caption{FSTD and OSTD for infinite-length $\mathcal{S}_1$-constrained sequence.} \label{fig:OSTM_S1}
\vspace{-1.2em}
\end{figure}

%We shall consider the transitions in the corresponding OSTD. Each edge transits a sequence of ${\bf 0}^{t-1}1$ with weight $t$ in FSTD.
%Since the input for states $B$ and $C$ in FSTD is $1$, they will be only two states in OSTD. We obtain Fig.~\ref{subfig:OSTM_S1_c} considering the following $4$ cases:
%\begin{enumerate}
%    \item State $B$ does not have self-loop, so the new state $1$ does not have self-loop. 
%    \item State $C$ has a self-loop, so the new state $2$ has a self-loop with weight $1$. 
%    \item $B\to C$ is the only transition from state $B$ to $C$, so the weight for the transition from the new state $1$ to $2$ is $1$. 
%    \item The transitions from state $C$ to $B$ have length at least $3$, which is given by $C\to D\to A\to B$. Since state $A$ has a self-loop that always transits $0$, the weight on the edge from state $1$ to $2$ can be any $t\in\mathbb{Z}$ where $t\geq 3$.
%\end{enumerate}
%We also keep track of the probabilities associated with the weights in OSTD in Fig.~\ref{subfig:OSTM_S1_d}. It follows from~(\ref{eq:def_G_D}) that the OSTM of $\mathcal{S}_1$-constrained code is given by

We again obtain Fig.~\ref{subfig:OSTM_S1_c} by using only the labeled states and drawing all the possible transitions from one state to another. The edges are labeled by run-length(s) $t_i$, indicating a transition corresponding to a sequence ${\bf 0}^{t_i-1}1$. We continue to keep track of the probabilities associated with the edge labels in the OSTD in Fig.~\ref{subfig:OSTM_S1_d}. It then follows from~(\ref{eq:def_G_D}) that the OSTM of infinite-length $\mathcal{S}_1$-constrained code is given by
\begin{align}
    {\bf{G}}(D) = \begin{bmatrix}
    0 & D\\
    \alpha & \frac{1}{2}D
    \end{bmatrix},
\end{align}
where
\vspace{-0.15em}\begin{align}
    \alpha = \sum_{k=0}^\infty \left(\frac{1}{2}\right)^{2+k}D^{3+k} = \frac{D^3}{2(2-D)}.
\end{align}
\end{example}

\begin{result}
\label{res:OSTM_S}
According to our derivation by induction, the OSTM for any $\mathcal{S}_x$-constrained code is then
\vspace{-0.4em}\begin{align}
    {\bf{G}}(D)=\begin{bmatrix} 
        \begin{matrix}
        0 \\ \vdots \\ 0
        \end{matrix} & D{\bf{I}}_x \\
        \alpha & \begin{matrix}
        0 & \cdots & 0 & \frac{1}{2}D
        \end{matrix}
        \end{bmatrix},
\end{align}
%  \begin{pspicture}
%  $ \begin{pmatrix}
%     \pnode[-0.3ex, 1.5ex]{A}
%     a_{11} & a_{12} & a_{13} & b_{11} & b_{12} \\
%     a_{21} & a_{22} & a_{23} & b_{21} & b_{22} \\
%     c_{11} & c_{12} &
%     c_{13} \pnode[1.4ex, -0.8ex]{C} & d_{11} & d_{12} \\
%     c_{21} & c_{22} & c_{23} & d_{21} & d_{22} \\
%     c_{31} & c_{32} & c_{33} & d_{31} & d_{32}\pnode[0.2ex, -0.6ex]{D}\,
%   \end{pmatrix} $
% \psset{linecolor = red}
% \psframe(A)(C)
% \psset{dimen = inner}\psframe(C)(D)
% \end{pspicture}
where 
\vspace{-0.15em}\begin{align}
    \alpha = \sum_{k=0}^{\infty}\left(\frac{1}{2}\right)^{2+k}D^{2+x+k}=\frac{D^{2+x}}{2(2-D)}.
\end{align}
\end{result}

 \begin{figure}
\centering
\includegraphics[width=10cm]{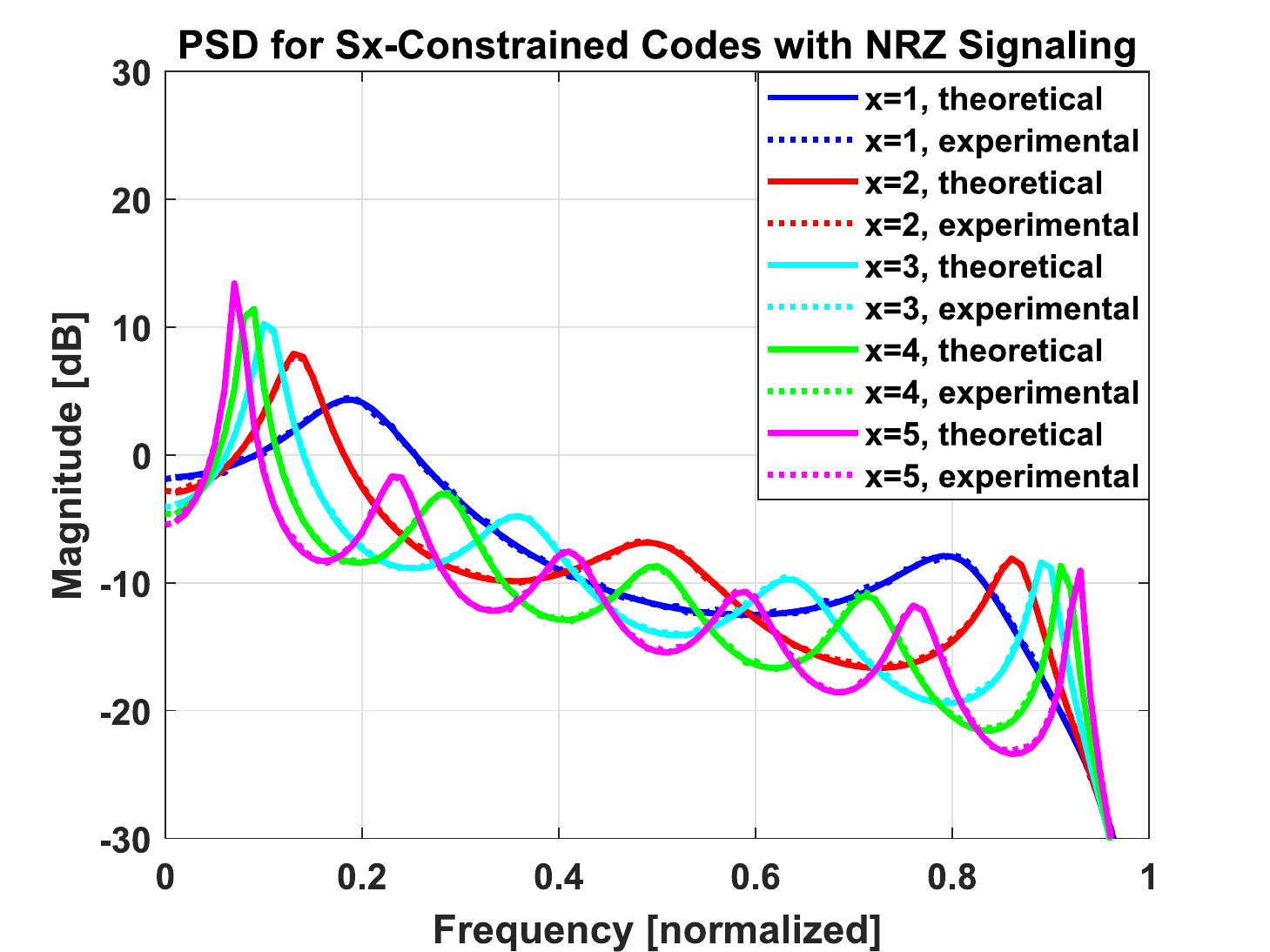}
\caption{PSDs for infinite-length $\mathcal{S}_x$-constrained codes with NRZ signaling.}\label{fig:Sx_PSD}
\end{figure} 

We then generate the PSD for an infinite-length $\mathcal{S}_x$-constrained code following~(\ref{eq:thm_Sx(D)}),~(\ref{eq:Sy(D)}),~(\ref{eq:Sw(f)}), and~(\ref{eq:Sw(0)}). Fig.~\ref{fig:Sx_PSD} shows the PSDs for these codes with $x \in \{1, 2, \dots, 5\}$.

It is clear from both Fig.~\ref{fig:Ax_PSD} and Fig.~\ref{fig:Sx_PSD} that the theoretical PSD plots match perfectly the experimental (Monte-Carlo) ones.

It should be noted that it is possible to use the exact method described in \cite{gall_psd} without the modifications described in this paper to find the PSD of these infinite-length constrained codes. For further details, see Appendix~\ref{ap:A}.

%%%%%%%%%%%%%%%%%%%%%%%%%%%%%%%%%%%%%
\section{OSTM Generation Method for Finite-Length Codes}\label{sec_finiteFSTD}

When the constraints adopted in $\mathcal{A}_x$-constrained and $\mathcal{S}_x$-constrained codes are imposed on a finite-length codebook, the FSTD and thus the OSTM become more complicated due to the possible transitions being dependent on the bit location within the codeword and its bridging pattern. In these finite-length constrained codes, preventing patterns constrains the beginning bits and the ending bits of a codeword in a way that differs from how it constrains the middle bits. Observe that the length of the shortest forbidden pattern is $3$. Another characteristic of finite-length codes is that they require bridging patterns. Given any two consecutive codewords, bridging patterns are bits/symbols appearing between the codewords to prevent forbidden patterns from occurring on the transition from any codeword to the following codeword.

The method for generating the FSTD/OSTM that is based on~\cite{gall_psd} and described in Section~\ref{sec_generalNRZ} provides the backbone to a generalized method for finding the FSTD/OSTM for finite-length codes constrained by the separation of transitions such as LOCO codes and A-LOCO codes described in detail later. The steps of our generalized method are:

\begin{enumerate}
    \item Derive FSTD states by considering a grid of states where each column represents a bit position within the codeword and its bridging pattern, and each row represents a possible bit sequence consisting of the previous $x$ bits and the current bit. This results in a maximum of $(m+x)2^{x+1}$ states, since there are $m+x$ bit positions and $2^{x+1}$ possible $x+1$ binary sequences. However, we will see that forbidden patterns result in some states being unneeded. Note that $x$ and $m$ are parameters of LOCO and A-LOCO codes. Here, $x$ defines the set of forbidden patterns as seen in $\mathcal{S}_{x}$-constrained and $\mathcal{A}_{x}$-constrained codes, while $m$ is the length of each codeword without its bridging.
    \item Connect states in each column to states in the next column with reference to allowed patterns according to the codebook constraint. These are always single bit transitions.
    \item Note the probabilities of each transition in the FSTD using the codebook. For convenience, all codewords in the codebook are assumed to be available for usage.
    \item Discard states that are never achieved, i.e., never connected to other states.
    \item Label states that result from the most recent bit being a $1$.
    \item If desirable, combine states that have the same exiting paths and probabilities.
    \item Develop the OSTM ${\bf{G}}(D)$ associated with this final FSTD by using the edge probabilities and run lengths, as done in Section~\ref{sec_generalNRZ}.
    \item If desirable for a class of codes, look for a pattern in the OSTMs of codes with specific parameters that leads to a general solution for the OSTM of that class of codes.
    \item Follow Theorem~\ref{thm:1} for obtaining the final PSD of the code.
\end{enumerate}

If examined closely, this method provides key insights regarding how the OSTM of these codes grows with respect to $x$ and $m$. The maximum number of states generated, assuming that they all can be achieved and therefore not discarded, is $(m+x)2^{x+1}$. The maximum number of labeled states is $\frac{(m+x)2^{x+1}}{2}=(m+x)2^{x}$, since only states resulting from a $1$ are labeled. Because the OSTM is a square matrix whose size is defined by the maximum number of labeled states, the OSTM is at maximum size an $(m+x)2^{x} \times (m+x)2^{x}$ matrix. This means that as the length of the codewords $m$ grows, the matrix dimension grows linearly. However, as the transition-separation constraint variable $x$ grows, the matrix dimension grows exponentially. In practice, $x$ tends to stay small, within $\{1, 2, 3\}$ typically, while $m$ can be large depending on the specific rate requirement \cite{ahh_loco, ahh_aloco}, so these trends are favorable. 

As discussed later in this paper, some codes do not result in FSTDs with this bound on the maximum number of labeled states, specifically clocked LOCO and clocked A-LOCO codes \cite{ahh_loco, ahh_qaloco}. Self-clocking is an important property for various line codes, as it enables the receiver to obtain the clock of the transmitter from the signal. Thus, in Appendix~\ref{ap:B}, we give the definitions of self-clocked A-LOCO and LOCO codes and, as an alternative, describe a general but slower algorithmic approach to construct the OSTM using breadth first search (BFS). Additionally, it is often the case that as the OSTM size grows, the OSTM becomes more sparse. This is particularly true in the upcoming examples of LOCO and A-LOCO codes.

We apply this method to study the spectra of A-LOCO and LOCO codes in Section~\ref{sec_finite}. We provide the general OSTM solution for any A-LOCO and LOCO code with standard, optimal bridging, and we also show how to address other finite-length challenges.

%%%%%%%%%%%%%%%%%%%%%%%%%%%%%%%%%%%%%
\section{Closed-Form PSD of Finite-Length Constrained Codes}\label{sec_finite}

Before calculating the PSD for a finite-length constrained code, we begin by referencing Fig.~\ref{fig:Ax_PSD}, the PSDs of infinite-length $\mathcal{A}_x$-constrained codes. A spike (delta function) at frequency $0$ (DC) is seen. This implies that there is some discrete amount of power at DC, which intuitively makes sense because $p(1)$ is not $\frac{1}{2}$ as in $\mathcal{S}_x$-constrained codes. Instead, $p(1)=\frac{2}{x+4}$ for $\mathcal{A}_x$-constrained codes. This implies that the level-based sequence $\{Y_n\}$ has some DC offset. Mathematically, this means that there is a periodic component in the auto-correlation function associated with $\{Y_n\}$. In this specific case of infinite-length $\mathcal{A}_x$-constrained codes, this component is a constant $= (\mathbb{E}[Y_n])^2 = \frac{x^2}{(x+4)^2}$, which results in a delta function at DC when calculating the spectrum.

Besides for the case of having discrete power at DC, it is possible for a code to have discrete power at other frequencies. This occurs when the code sequence process is cyclostationary, and the expected value at any stream position is not fixed at $0$ after signaling. In this case, the auto-correlation function has a periodic component that is not a constant.

\subsection{Discussion on the Cyclostationarity of Finite-Length Codes}
Time-synchronous pulse amplitude modulated signals are known to be cyclostationary with a period equal to the channel bit interval. A stream of $m$-length codewords exhibits higher level cyclostationarity with a period equal to $m+x$ channel bit intervals when bridging patterns are designed to be $x$-bits long. To account for these cyclic statistics in the PSD, phase-averaging is performed on the auto-correlation function over $m+x$ bits (the higher level cyclostationarity). As mentioned, it is cyclostationarity along with non-zero averages for $\{Y_n\}$ that cause the PSD of a finite-length code to have a discrete component and a continuous component. 

The continuous component and stationary statistics are obtained by considering the aperiodic component of the phase-averaged auto-correlation function. This is equivalent to considering the auto-covariance function rather than the auto-correlation function. The periodic component of the auto-correlation function then results in a discrete component in the PSD. This is seen as concentrated power at integer multiples of the codeword normalized frequency, $\frac{1}{m+x}$, including power at DC. In terms of data transmission, the discrete component is typically considered a somewhat undesirable use of power. However, these spectral lines can be used to extract timing or position information from the sequence in some applications \cite{immink_book}.

A general expression for the (average) PSD function of a cyclostationary code after NRZ signaling, i.e., of its write process $W(t)$ obtained from $\{Y_n\}$, is
\begin{align}
    S_W(f) = S_W^\textup{c}(f)+S_W^\textup{d}(f),
\end{align}

where $S_W^\textup{c}(f)$ and $S_W^\textup{d}(f)$ are the continuous and discrete PSD components, respectively. For a sequence of bridged codewords $\{X_n\}$, the phase-averaged auto-correlation function can be written as
\vspace{-0.1em}\begin{align}\label{eq:phase_av_autocor}
    \mathbb{R}_X(k) = \overline{\mathbb{E}} [X_\ell X_{\ell+k}] = \frac{1}{m+x} \sum_{\ell=0}^{m+x-1} \mathbb{E} [X_\ell X_{\ell+k}].
\end{align}
Averaging $\ell$ over the $m+x$ period is what makes the auto-correlation function phase-averaged. 

For a code with codeword length $m$ and bridging length $x$, the purely aperiodic component of the auto-correlation function used to calculate the continuous component of the PSD only exists for $k=0,\pm 1, \pm 2, \dots , \pm (m+x-1)$. Since the periodic component of the auto-correlation function is cyclic with intervals equal to integer multiples of $m+x$, the coefficients of the second period in the auto-correlation function can be subtracted from the overall auto-correlation coefficients within the first period to determine $\mathbb{R}_X^{\text{aperiodic}}(k)$: 
\begin{align}
    \mathbb{R}_X^{\text{aperiodic}}(k) =  
    \begin{cases}
        \mathbb{R}_X(k) - \mathbb{R}_X(k+m+x), & 0\leq k<m+x, \\
        \mathbb{R}_X(k) - \mathbb{R}_X(k-m-x), & -m-x<k<0, \\
        0, & \text{otherwise}.
    \end{cases}
\end{align}
It then follows that the periodic component of the auto-correlation function is
\begin{align}
    \mathbb{R}_X^{\text{periodic}}(k) =  \mathbb{R}_X(k)-\mathbb{R}_X^{\text{aperiodic}}(k).
\end{align}
All the characteristics regarding periodicity/aperiodicity for the process $\{Y_n\}$ are inherited from the process $\{X_n\}$. Thus, the analysis also extends to $\{Y_n\}$.

It is important to understand that Theorem~\ref{thm:1} and the discussion following it in Section~\ref{sec_generalNRZ} can be used to theoretically generate the continuous component of the PSD of a cyclostationary sequence. First, the process of using the OSTM ${\bf{G}}(D)$ to derive the continuous component of the PSD $S_X(D)$ inherently takes care of the phase-averaging for the code binary sequence $\{X_n\}$. Second, recall (\ref{eq:exp_y}), and now average the expectation over the period covering a codeword plus its bridging pattern in order to phase-average for $\mathbb{R}_Y(k)$ as shown in (\ref{eq:phase_av_autocor}): 
\begin{align}\label{eq:R_Y_cycol}
    \mathbb{R}_Y(k) &= \overline{\mathbb{E}}[Y_\ell Y_{\ell+k}] = \frac{1}{m+x} \sum_{\ell=0}^{m+x-1} \mathbb{E} [Y_\ell Y_{\ell+k}]\nonumber \\
    &= \frac{1}{m+x} \sum_{\ell=0}^{m+x-1} \mathbb{E} [(2X_\ell-1) (2X_{\ell+k}-1)]\nonumber \\
    &=\frac{4}{m+x}\sum_{\ell=0}^{m+x-1}\mathbb{E}[X_\ell X_{\ell+k}] - \frac{2}{m+x}\sum_{\ell=0}^{m+x-1}\mathbb{E}[X_\ell] - \frac{2}{m+x}\sum_{\ell=0}^{m+x-1}\mathbb{E}[X_{\ell+k}]+ 1\nonumber \\
    &=4\overline{\mathbb{E}}[X_{\ell} X_{\ell+k}] - 2\overline{\mathbb{E}}[X_\ell] - 2\overline{\mathbb{E}}[X_{\ell+k}]+ 1\nonumber \\
    &=4\overline{\mathbb{E}}[X_\ell X_{\ell+k}] - 4\overline{p}(1)+ 1 = 4\mathbb{R}_X(k) - 4\overline{p}(1)+ 1.
\end{align}
The overline notation here refers to averaging over a period of length $m+x$ via the variable $\ell$. Since $\overline{\mathbb{E}}[X_\ell]$ and $\overline{\mathbb{E}}[X_{\ell+k}]$ are averaged over the period of cyclostationarity, they are exactly the same. Their value is $\overline{p}(1)$, which is computed as shown in (\ref{eq:Phi_D_and_G_D}). More specifically, we see that they are the equilibrium probability. Using the overline notation leads to the same result as seen in (\ref{eq:exp_y}), therefore justifying the use of Theorem~\ref{thm:1} and its following discussion (the equations from (\ref{eq:exp_y}) to (\ref{eq:Sw(0)})) for cyclostationary finite-length constrained codes.

Assuming NRZ signaling and a rectangular pulse shape, the continuous component of the PSD of the code, i.e., of $W(t)$, can then be calculated as
\begin{align}
    S_W^\textup{c}(f) = \text{sinc} ^2(\pi f) S_Y^\textup{c}(e^{i2\pi f}) = \text{sinc} ^2(\pi f) \left [ \sum_{k=-m-x+1}^{m+x-1} \mathbb{R}_Y^{\text{aperiodic}}(k)D^{k} \right ]_{D = e^{i2\pi f}},
\end{align}
where $S_Y^\textup{c}(\cdot)$ is derived from $S_X^\textup{c}(\cdot)$ (both are the continuous PSD components) the same way described in (\ref{eq:Sy(D)}) and further justified by (\ref{eq:R_Y_cycol}).

Using Fourier analysis, the discrete component of the PSD then follows as
\begin{align}\label{eq:discrete_PSD}
    S_W^\textup{d}(f) = \text{sinc} ^2(\pi f) \left(\frac{a_0}{2} \delta(f) +\sum_{n=1}^\infty \frac{a_n}{2}\left(\delta\left(f-\frac{n}{m+x}\right)+\delta\left(f+\frac{n}{m+x}\right)\right)\right),
\end{align}
\begin{align*}
    \text{where } a_0=\frac{2}{m+x} \sum_{k=0}^{m+x-1} \mathbb{R}_Y^{\text{periodic}}(k)
\end{align*}
\begin{align*}
    \text{and } a_n=\frac{2}{m+x} \sum_{k=0}^{m+x-1} \mathbb{R}_Y^{\text{periodic}}(k) \cos{\left(2\pi\frac{n}{m+x}k\right)}.
\end{align*}
The $\delta(.)$ is the dirac delta function, and it occurs at integer multiples of the sequence frequency, $0$ and $\pm \frac{n}{m+x}$, where $n\in\mathbb{N}^+$.
% Let $\mathbf{f}$ be the row vector
%
%\begin{align}
%    \mathbf{f} = (e^{j2 \pi f}, e^{j4 \pi f}, \dots, e^{j2n \pi f})
%\end{align}
%
%Both \cite{Biglieri} and \cite{Cariolaro} provide a matrix %representation of the correlations of block-coded sequences. In %context of this paper, the blocks of $n$ bits are of length $m+x$.
%
%\begin{align}
%    R_k = \mathbb{E} [x_t^\mathsf{T} x_{t+k}], k=0,\pm 1, \pm 2,\dots,
%\end{align}
%
%To determine both the continuous and discrete parts of the power spectral density, the sequence of correlation matrices ($k=0,1,2,\dots$) and its limit ($R_\infty$) must be calculated.
%
%\begin{align}
%    S_{c}(f) = \frac{1}{m+x}\mathbf{f}(R_0-R_\infty)\mathbf{f}^{*}+\frac{2}{m+x}Re\sum_{k=1}^{\infty}\mathbf{f}(R_k-R_\infty)\mathbf{f}^{*}e^{-i2\pi k f (m+x)}
%\end{align}
%
%\begin{align}
%    S_{d}(f) = \frac{1}{(m+x)^{2}}\mathbf{f}R_\infty \mathbf{f}^{*}
%\end{align}
Further details on calculating these components can be found in \cite{immink_book}. By calculating the phase-averaged auto-correlation function using all possible codewords in the codebook or using many random streams of the desired code, an experimental solution for the PSD of the cyclostationary process was generated. This Monte-Carlo type solution is what we used to verify the theoretical results.

At frequencies where a discrete component is present, the OSTM used in the theoretical calculation becomes singular, causing inaccuracies in PSD values when used in (\ref{eq:thm_Sx(D)}) of Theorem~\ref{thm:1}. These points on the PSD can be calculated separately. We first note that
\begin{align}
    \mathbb{R}_X^{\text{periodic}}(k) = 
    \frac{1}{m+x} \sum_{\ell=0}^{m+x-1} \mathbb{E} [X_\ell] \mathbb{E} [X_{\ell+\tilde{k}}],
     \text{ where } \tilde{k}=\text{mod}(|k|,m+x)+m+x,
\end{align}
which requires only $m+x$ calculations. The same applies for $\mathbb{R}_Y^{\text{periodic}}(k)$. Then, we use (\ref{eq:discrete_PSD}).

\subsection{General Solution for the PSD of any A-LOCO Code}\label{subsec_A-LOCO}
We first consider A-LOCO codes, adopting their definition from \cite{ahh_aloco}. 

\begin{definition}%~\cite{ahh_aloco}
An A-LOCO code $\mathcal{AC}_{m,x}$, with parameters $m \geq 1$ and $x \geq 1$, is defined via the following properties:
\begin{enumerate}
    \item Codewords in $\mathcal{AC}_{m,x}$ are binary and of length $m$.
    \item Codewords in $\mathcal{AC}_{m,x}$ are ordered lexicographically.
    \item Any pattern in the asymmetric set $\mathcal{A}_{x}$ (see Definition~\ref{def:Ax_set}) does not appear in any codeword $\bold{c}$ in $\mathcal{AC}_{m,x}$.
    \item The code $\mathcal{AC}_{m,x}$ contains all the codewords satisfying the previous three properties.
\end{enumerate}
Lexicographic ordering of codewords means that they are ordered in an ascending manner following the rule $0 < 1$ for any bit, and the bit significance reduces from left to right.
\end{definition}
As was done with infinite-length $\mathcal{A}_x$-constrained codes, we have categorized A-LOCO codes into the family of codes constrained by transition separation. We adopt a simple bridging pattern of $\bold{0}^x$ unless the ending bit of a codeword and the beginning bit of the next codeword are both $1$'s, at which point the bridging pattern is $\bold{1}^x$.

\begin{example}
We illustrate our method for obtaining the PSD of A-LOCO codes in this example. Consider a code whose set of forbidden patterns is $\{101\}$ and all codewords are of length $4$, i.e., $x=1$ and $m=4$. We adopt the aforementioned bridging. This is equivalently $\mathcal{AC}_{4,1}$ with its standard method of bridging which is optimal in terms of bit protection, easy to implement, and requires the minimum number of added bits \cite{ahh_aloco}.

In accordance with our method, the initial FSTD is given in Fig.~\ref{fig:A-LOCO_Init}. Then, if unused states are deleted, only states resulting from the most recent bit being a $1$ are labeled in gray, and states are combined when desirable, the final FSTD becomes Fig.~\ref{fig:A-LOCO_Fin}.

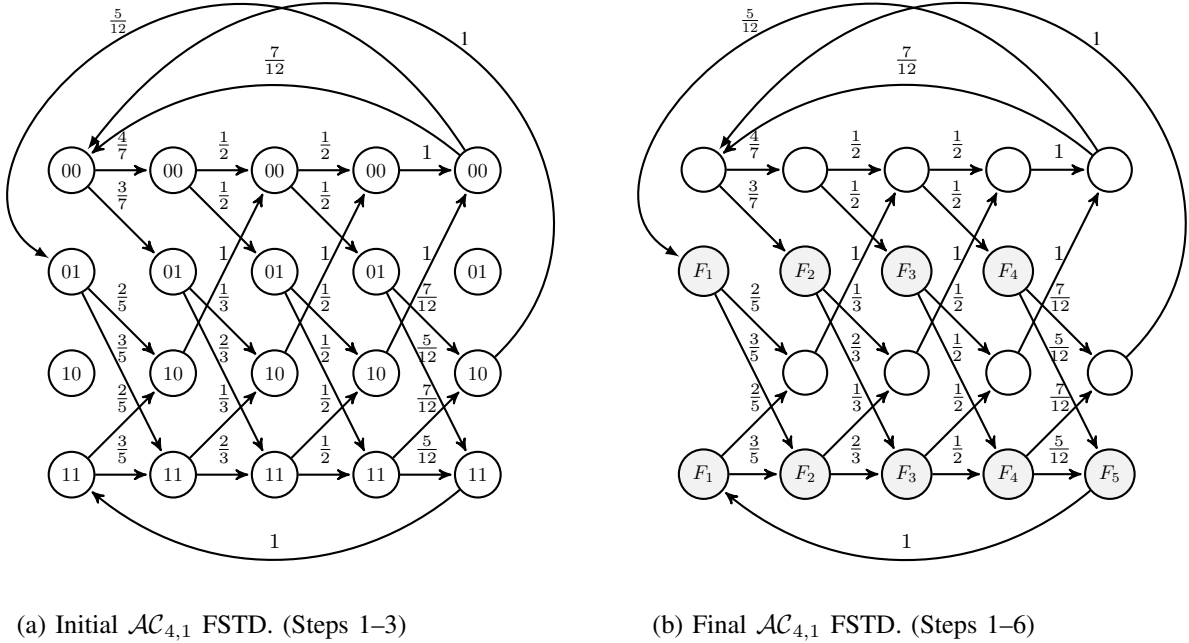
\begin{figure}
\vspace{-6.0em}
\hspace{-2.2em}\begin{subfigure}[b]{.5\textwidth}
    \centering
    \begin{tikzpicture}[->,>=stealth',shorten >=1pt,auto,node distance=1.8cm, thick, scale=0.75, every node/.append style={transform shape}, main node/.style={circle,draw,font=\sffamily\small\bfseries},every edge/.append style={nodes={font=\sffamily\scriptsize}},every path/.append style={nodes={font=\sffamily\scriptsize}}]
    \node[main node] (1) {$00$};
    \node[main node] (2) [below of=1] {$01$};
    \node[main node] (3) [below of=2]{$10$};
    \node[main node] (4) [below of=3]{$11$};
    \node[main node] (5) [right of=1] {$00$};
    \node[main node] (6) [below of=5] {$01$};
    \node[main node] (7) [below of=6]{$10$};
    \node[main node] (8) [below of=7]{$11$};
    \node[main node] (9) [right of=5] {$00$};
    \node[main node] (10) [below of=9] {$01$};
    \node[main node] (11) [below of=10]{$10$};
    \node[main node] (12) [below of=11]{$11$};
    \node[main node] (13) [right of=9] {$00$};
    \node[main node] (14) [below of=13]{$01$};
    \node[main node] (15) [below of=14]{$10$};
    \node[main node] (16) [below of=15]{$11$};
    \node[main node] (17) [right of=13] {$00$};
    \node[main node] (18) [below of=17] {$01$};
    \node[main node] (19) [below of=18]{$10$};
    \node[main node] (20) [below of=19]{$11$};
    \path[every node/.style={font=\sffamily\small}]
        (1) edge node [above] {$\frac{4}{7}$} (5)
            edge node[above] {$\frac{3}{7}$} (6)
        (2) edge node [above] {$\frac{2}{5}$} (7)
            edge node[above] {$\frac{3}{5}$} (8)  
        (4) edge node [above] {$\frac{2}{5}$} (7)
            edge node[above] {$\frac{3}{5}$} (8)
        (5) edge node [above] {$\frac{1}{2}$} (9)
            edge node[above] {$\frac{1}{2}$} (10)
        (6) edge node [above] {$\frac{1}{3}$} (11)
            edge node[above] {$\frac{2}{3}$} (12) 
        (7) edge node [above] {$1$} (9)
        (8) edge node [above] {$\frac{1}{3}$} (11)
            edge node[above] {$\frac{2}{3}$} (12)
        (9) edge node [above] {$\frac{1}{2}$} (13)
            edge node[above] {$\frac{1}{2}$} (14)
        (10) edge node [above] {$\frac{1}{2}$} (15)
            edge node[above] {$\frac{1}{2}$} (16) 
        (11) edge node [above] {$1$} (13)
        (12) edge node [above] {$\frac{1}{2}$} (15)
            edge node[above] {$\frac{1}{2}$} (16) 
        (13) edge node [above] {$1$} (17)
        (14) edge node [above] {$\frac{7}{12}$} (19)
            edge node[above] {$\frac{5}{12}$} (20) 
        (15) edge node [above] {$1$} (17)
        (16) edge node [above] {$\frac{7}{12}$} (19)
            edge node[above] {$\frac{5}{12}$} (20) 
        (17) edge [bend right=40] node [above] {$\frac{7}{12}$} (1)
        (20) edge [bend left=40] node [above] {$1$} (4);
    \path [line] (19) .. controls (11,0) and (5,6.5) .. (1) node[transition,pos=0.5,above] {1};
    \path [line] (17) .. controls (3,6.5) and (-3,0) .. (2) node[transition,pos=0.5,above] {$\frac{5}{12}$};
    \end{tikzpicture}
    \caption{Initial $\mathcal{AC}_{4,1}$ FSTD. (Steps 1--3)}\label{fig:A-LOCO_Init}
\end{subfigure}
\begin{subfigure}[b]{.5\textwidth}
    \centering
    \begin{tikzpicture}[->,>=stealth',shorten >=1pt,auto,node distance=1.8cm, thick, scale=0.75, every node/.append style={transform shape}, main node/.style={circle,draw,font=\sffamily\small\bfseries},every edge/.append style={nodes={font=\sffamily\scriptsize}},every path/.append style={nodes={font=\sffamily\scriptsize}}] 
    \node[main node] (1) {\textcolor{white}{$D$}};
    \node[main node] (2) [below of=1, fill=gray!10] {\small$F_1$};
    \node[main node] (4) [below of=3, fill=gray!10]{\small$F_1$};
    \node[main node] (5) [right of=1] {\textcolor{white}{$D$}};
    \node[main node] (6) [below of=5, fill=gray!10] {\small$F_2$};
    \node[main node] (7) [below of=6]{\textcolor{white}{$D$}};
    \node[main node] (8) [below of=7, fill=gray!10]{\small$F_2$};
    \node[main node] (9) [right of=5] {\textcolor{white}{$D$}};
    \node[main node] (10) [below of=9, fill=gray!10] {\small$F_3$};
    \node[main node] (11) [below of=10]{\textcolor{white}{$D$}};
    \node[main node] (12) [below of=11, fill=gray!10]{\small$F_3$};
    \node[main node] (13) [right of=9] {\textcolor{white}{$D$}};
    \node[main node] (14) [below of=13, fill=gray!10]{\small$F_4$};
    \node[main node] (15) [below of=14]{\textcolor{white}{$D$}};
    \node[main node] (16) [below of=15, fill=gray!10]{\small$F_4$};
    \node[main node] (17) [right of=13] {\textcolor{white}{$D$}};
    \node[main node] (19) [below of=18]{\textcolor{white}{$D$}};
    \node[main node] (20) [below of=19, fill=gray!10]{\small$F_5$};
    \path[every node/.style={font=\sffamily\small}]
        (1) edge node [above] {$\frac{4}{7}$} (5)
            edge node[above] {$\frac{3}{7}$} (6)
        (2) edge node [above] {$\frac{2}{5}$} (7)
            edge node[above] {$\frac{3}{5}$} (8)  
        (4) edge node [above] {$\frac{2}{5}$} (7)
            edge node[above] {$\frac{3}{5}$} (8)
        (5) edge node [above] {$\frac{1}{2}$} (9)
            edge node[above] {$\frac{1}{2}$} (10)
        (6) edge node [above] {$\frac{1}{3}$} (11)
            edge node[above] {$\frac{2}{3}$} (12) 
        (7) edge node [above] {$1$} (9)
        (8) edge node [above] {$\frac{1}{3}$} (11)
            edge node[above] {$\frac{2}{3}$} (12)
        (9) edge node [above] {$\frac{1}{2}$} (13)
            edge node[above] {$\frac{1}{2}$} (14)
        (10) edge node [above] {$\frac{1}{2}$} (15)
            edge node[above] {$\frac{1}{2}$} (16) 
        (11) edge node [above] {$1$} (13)
        (12) edge node [above] {$\frac{1}{2}$} (15)
            edge node[above] {$\frac{1}{2}$} (16) 
        (13) edge node [above] {$1$} (17)
        (14) edge node [above] {$\frac{7}{12}$} (19)
            edge node[above] {$\frac{5}{12}$} (20) 
        (15) edge node [above] {$1$} (17)
        (16) edge node [above] {$\frac{7}{12}$} (19)
            edge node[above] {$\frac{5}{12}$} (20) 
        (17) edge [bend right=40] node [above] {$\frac{7}{12}$} (1)
        (20) edge [bend left=40] node [above] {$1$} (4);
    \path [line] (19) .. controls (11,0) and (5,6.5) .. (1) node[transition,pos=0.5,above] {1};
    \path [line] (17) .. controls (3,6.5) and (-3,0) .. (2) node[transition,pos=0.5,above] {$\frac{5}{12}$};
    \end{tikzpicture}
    \caption{Final $\mathcal{AC}_{4,1}$ FSTD. (Steps 1--6)}\label{fig:A-LOCO_Fin}
\end{subfigure}
\caption{FSTD used for OSTM generation of an $\mathcal{AC}_{4,1}$ sequence.}
\end{figure}

\begin{figure}
\vspace{-1.5em}
\centering
\includegraphics[width=10cm]{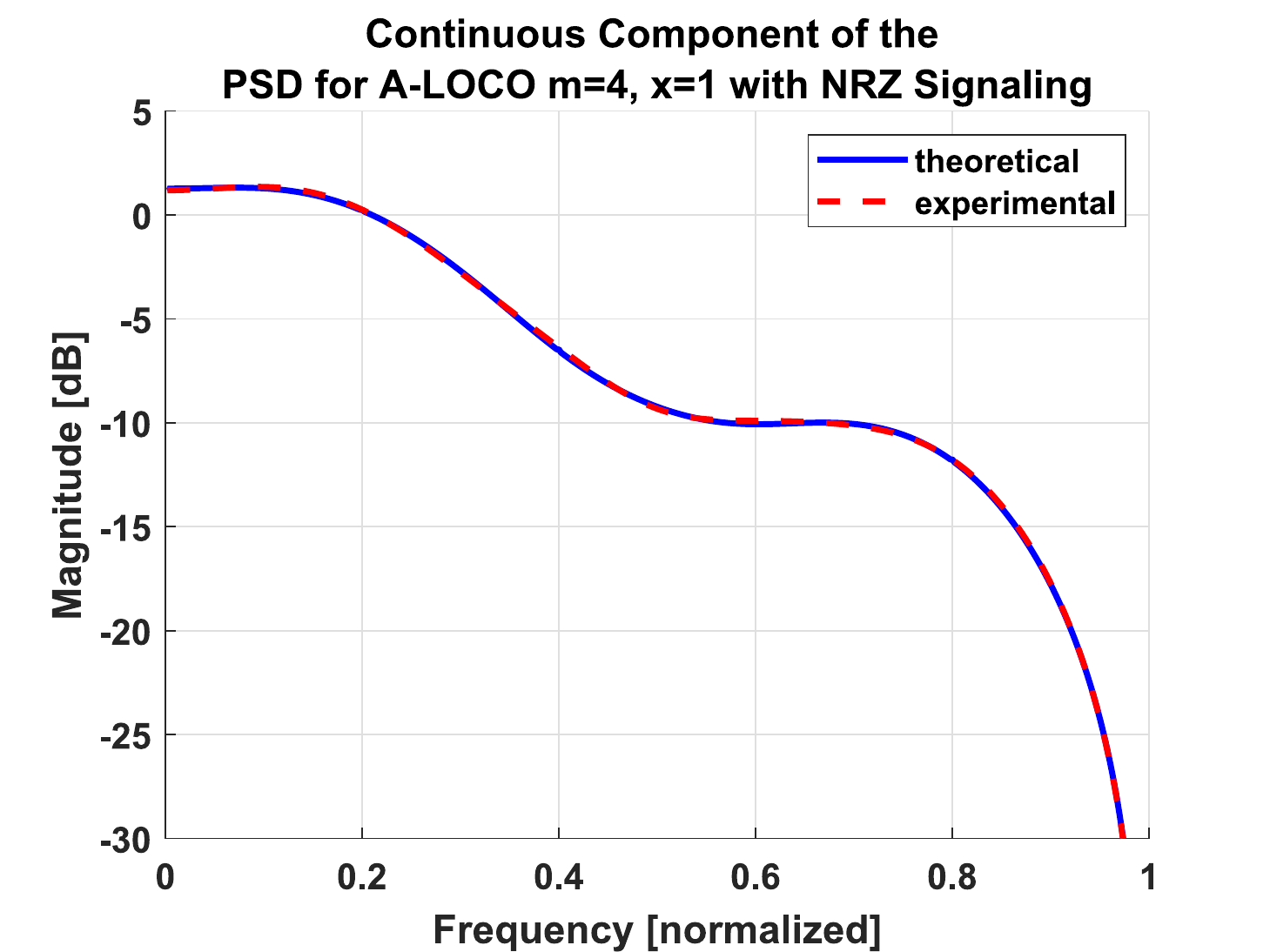}
\caption{The theoretical (OSTM method) versus experimental (Monte-Carlo) continuous component of the PSD for A-LOCO code $\mathcal{AC}_{4,1}$ with NRZ signaling.}\label{fig:cont_ALOCO}
\vspace{-2.0em}
\end{figure}

The resulting OSTM is then 
\begin{align}
\label{eq:G_D_A-LOCO_4_1}
    {\bf{G}}(D) = \begin{bmatrix}
    \beta_{1,5} & \frac{3}{5}D + \beta_{\frac{3}{5},6} & \beta_{\frac{2}{5},7} & \frac{1}{5}D^3 + \beta_{\frac{1}{5},8} & 0\\
    \beta_{\frac{5}{3},4} & \beta_{1,5} & \frac{2}{3}D + \beta_{\frac{2}{3},6} & \beta_{\frac{1}{3},7} & 0\\
    \beta_{\frac{5}{2},3} & \beta_{\frac{3}{2},4} & \beta_{1,5} & \frac{1}{2}D + \beta_{\frac{1}{2},6} & 0\\
    \beta_{\frac{5}{12},7} & \beta_{3,3} & \beta_{2,4} & \beta_{1,5} & \frac{5}{12}D\\
    D & 0 & 0 & 0 & 0\\
    \end{bmatrix},
\end{align}
where $\beta_{a,b} = \sum_{k=1}^\infty a\left( \frac{1}{12} \right) ^k D^{b+5(k-1)} = \frac{aD^{b}}{12-D^5}$.

Once the OSTM is found, we use Theorem~\ref{thm:1} and its following discussion to calculate the continuous component of the PSD for $\mathcal{AC}_{4,1}$ theoretically, see Fig.~\ref{fig:cont_ALOCO}.

The discrete component is then calculated using (\ref{eq:discrete_PSD}) and is shown in Fig.~\ref{fig:discrete_ALOCO}. For this A-LOCO code, the amplitudes in one period of the periodic component of the auto-correlation function were found to be $\mathbb{R}_Y^{\textup{periodic}}(0)=0.0964 \text{, } \mathbb{R}_Y^{\textup{periodic}}(1)=0.0436 \text{, } \mathbb{R}_Y^{\textup{periodic}}(2)=0.0056 \text{, } \mathbb{R}_Y^{\textup{periodic}}(3)=0.0056 \text{, and } \mathbb{R}_Y^{\textup{periodic}}(4)=0.0436$.

\begin{figure}
\centering
\includegraphics[width=10cm]{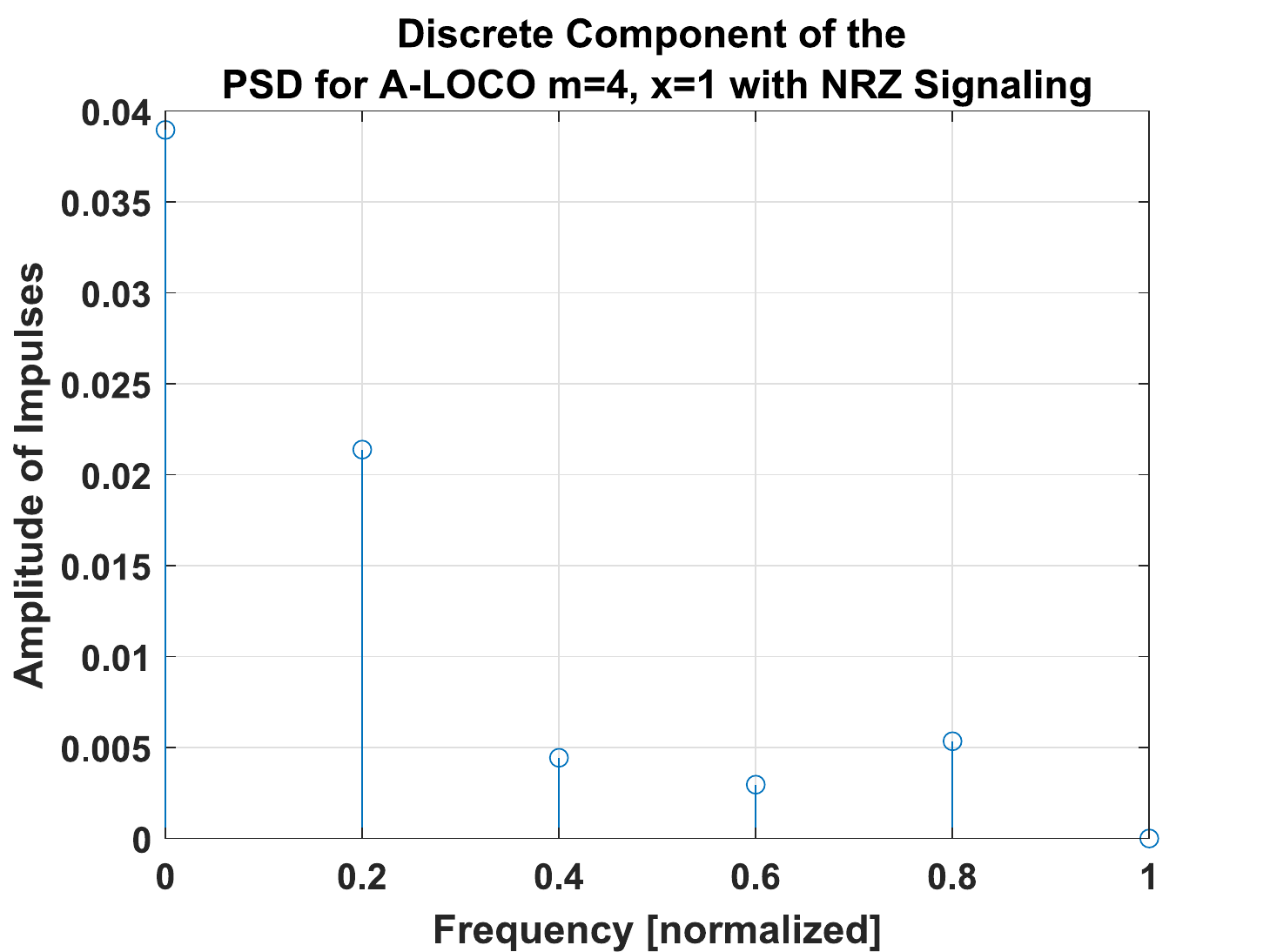}
\caption{The calculated discrete component of the PSD for A-LOCO code $\mathcal{AC}_{4,1}$ with NRZ signaling. Since the discrete component consists of delta functions, the result is displayed as the amplitude of the impulses (describing the area under the impulses) at the frequencies indicated.}\label{fig:discrete_ALOCO}
\vspace{-0.3em}
\end{figure}

\end{example}

\begin{result}
The general OSTM solution for any A-LOCO code $\mathcal{AC}_{m,x}$ defined by the parameters $m$ and $x$, and using $x$-bit bridging is the $m+x$ square matrix
\begin{align}
    {\bf{G}}(D) = \begin{bmatrix}
    {\bf{A}}_{m \times m} & \begin{matrix}
    0 & \text{\textcolor{white}{D}}{\bf{0}}\text{\textcolor{white}{D.}} \\
    \vdots & \text{\textcolor{white}{D}}\vdots\text{\textcolor{white}{D.}} \\
    0 & \text{\textcolor{white}{D}}\vdots\text{\textcolor{white}{D.}} \\
    \zeta D & \text{\textcolor{white}{D.}}{\bf{0}}\text{\textcolor{white}{D.}}
    \end{matrix}
    \\
    \begin{matrix}
    \bold{0} & \cdots & \cdots \\
    D & 0 & \cdots \\
    \end{matrix}
    &
    \begin{matrix}
    \bold{0} & D {\bf{I}}_{x-1} \\
    \cdots & \bold{0}\\
    \end{matrix}\\
    \end{bmatrix},
\end{align}
where $\zeta = \frac{N_2(m,x)+N_3(m,x)}{N(m,x)}$. The notation $N(m,x)$, $N_2(m,x)$, and $N_3(m,x)$ refers to the cardinality of $\mathcal{AC}_{m,x}$, the cardinality of Group 2 in $\mathcal{AC}_{m,x}$ (the number of codewords starting with $11$ from the left), and the cardinality of Group 3 in $\mathcal{AC}_{m,x}$ (the number of codewords starting with $1\bold{0}^{x+1}$ from the left), respectively, adopted from \cite{ahh_aloco}.

The entries of the submatrix ${\bf{A}}_{m \times m}$, which are $\bold{A}_{i,j}$, $1 \leq i,j \leq m$, can then be found using the following rules: 
\begin{itemize}
  \item If $j=i$, then  ${\bf{A}}_{i,j}=\beta_{1,m+x}$.
  \item If $j=i+1$ or $j>i+1+x$, then ${\bf{A}}_{i,j}=\lambda_{m+1-i,j-i-1}D^{j-i}+\beta_{\lambda_{m+1-i,j-i-1},m-i+j+x}$.
  \item If $i+2\leq j \leq i+1+x$, then ${\bf{A}}_{i,j}=\beta_{\lambda_{m+1-i,j-i-1},m-i+j+x}$.
  \item If $j<i$, then ${\bf{A}}_{i,j}=\beta_{\frac{1}{\lambda_{m+1-j,i-j-1}},m-i+j+x}$.
  \item If $i=m$ and $j=1$, then ${\bf{A}}_{i,j}=\beta_{\zeta,m+2x+1}$.
\end{itemize}

In those rules, 
\begin{align}
    \beta_{a,b} = \sum_{k=1}^\infty a\left( \frac{1}{N(m,x)} \right) ^k D^{b+(m+x)(k-1)} = \frac{a D^b}{N(m,x)-D^{m+x}},
\end{align}
\begin{align}
    \text{and } \lambda_{d,g}=\prod_{k=0}^{g}(1-\alpha_{d-k}),
\end{align}
\begin{align}
    \text{where } \alpha_c = \frac{N_3(c,x)}{N_2(c,x)+N_3(c,x)}.
\end{align}
\end{result}

\begin{figure}
    \begin{subfigure}[b]{.5\textwidth}
        \centering
        \includegraphics[width=3.25in]{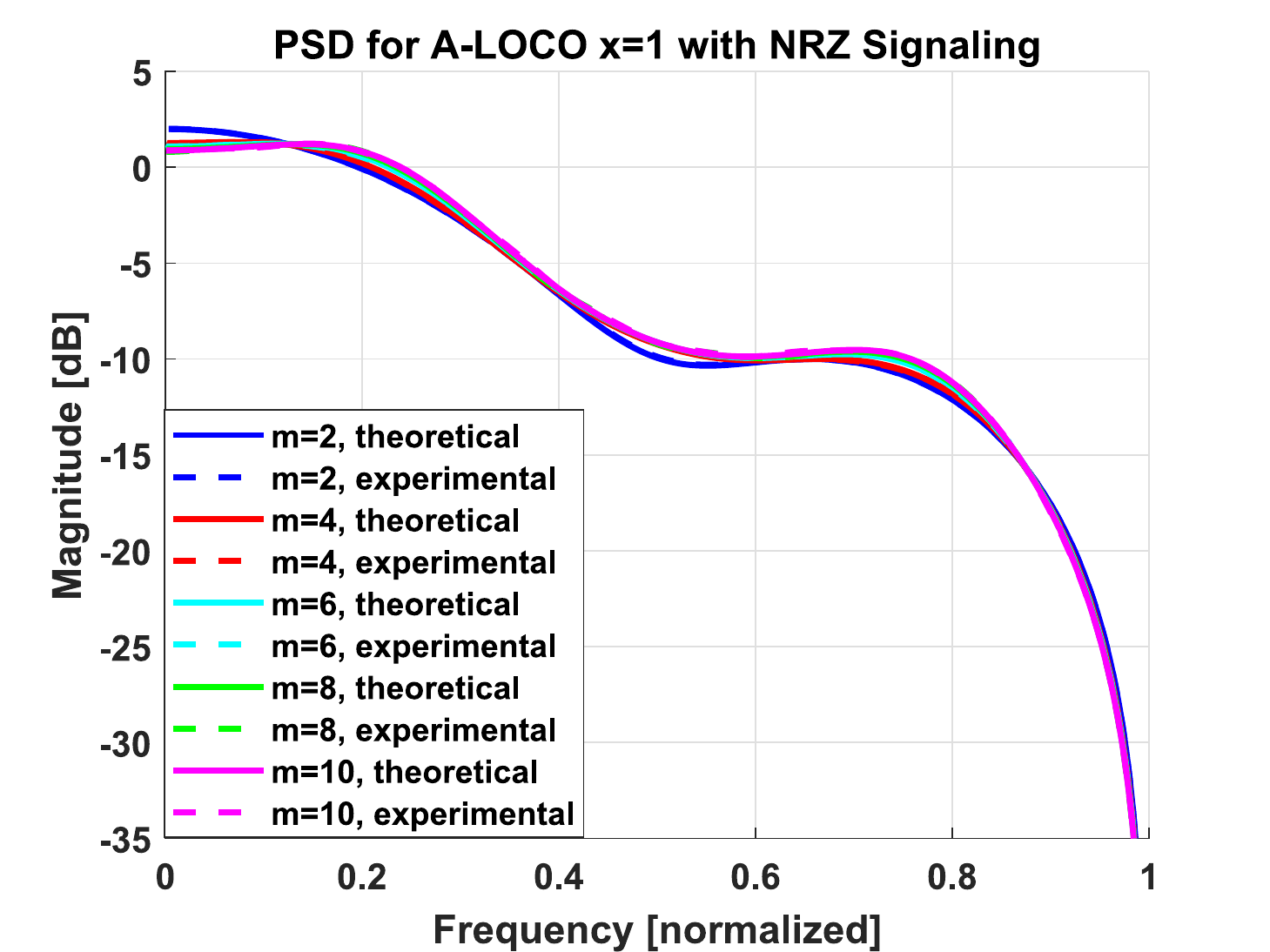}
    \end{subfigure}
    \begin{subfigure}[b]{.5\textwidth}
        \centering
        \includegraphics[width=3.25in]{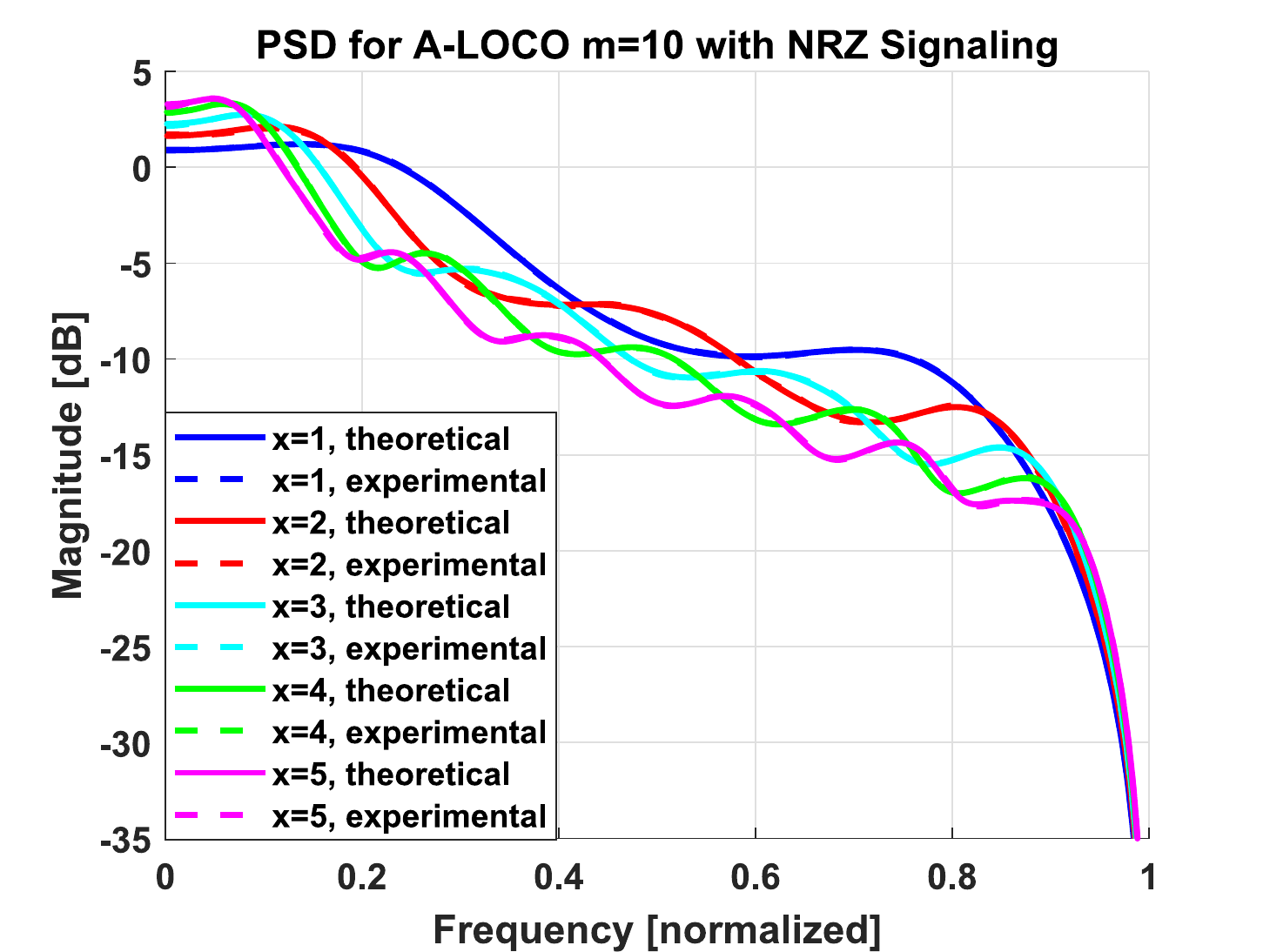}
    \end{subfigure}
    \caption{Comparisons of the continuous component of various $\mathcal{AC}_{m,x}$ spectra.}
    \label{fig:A-LOCOspectra}
\end{figure}

Since A-LOCO codes are asymmetric, there is an expected amount of discrete power at frequency $0$ (and other frequencies as well) since $0$'s are more likely to occur than $1$'s. Moreover, as $x$ becomes smaller compared with $m$, the continuous PSD at frequency $0$ decreases. These and other characteristics of A-LOCO spectra, such as the number of spectral peaks increasing as $x$ increases, are demonstrated in Fig.~\ref{fig:A-LOCOspectra}.

One standard performance measurement of a signal is the $3$dB bandwidth defined as twice the distance from frequency $0$ to the point at which the magnitude becomes $3$ dB lower than the magnitude at frequency $0$. Due to the spike at frequency $0$ in asymmetric codes such as A-LOCO codes, the magnitude at frequency $0$ here is considered to be the limit of the magnitude as frequency approaches $0$, which is the magnitude of the continuous PSD at frequency $0$. For the examples in Fig.~\ref{fig:A-LOCOspectra}, the $3$dB bandwidths are presented in Table~\ref{tb:aloco}. Here, the $3$dB bandwidth decreases for the same $m$ as $x$ increases since fast transitions in $W(t)$ become more restricted.

\begin{table}
\centering
 \caption{$3$dB Bandwidth for Various A-LOCO Codes}\label{tb:aloco}
 \begin{tabular}{||p{3.0cm}||c|c|c|c|c|c|c|c|c||} 
 \hline
 \multicolumn{10}{||c||}{A-LOCO $3$dB Bandwidth Measurements} \\
 \hline\hline
 $x$ [bits] & $1$ & $1$ & $1$ & $1$ & $1$ & $2$ & $3$ & $4$ & $5$  \\ 
 \hline
 $m$ [bits] & $2$ & $4$ & $6$ & $8$ & $10$ & $10$ & $10$ & $10$ & $10$ \\
 \hline
 $3$dB BW [normalized] & $0.480$ & $0.542$ & $0.577$ & $0.591$ & $0.596$ & $0.431$ & $0.334$ & $0.273$ & $0.231$ \\
 \hline
 \end{tabular}
 \end{table}

\subsection{General Solution for the PSD of any LOCO Code}\label{subsec_LOCO}
We next consider LOCO codes, adopting their definition from \cite{ahh_aloco}.

\begin{definition}%~\cite{ahh_loco}
A LOCO code $\mathcal{C}_{m,x}$, with parameters $m \geq 1$ and $x \geq 1$, is defined via the following properties:
\begin{enumerate}
    \item Codewords in $\mathcal{C}_{m,x}$ are binary and of length $m$.
    \item Codewords in $\mathcal{C}_{m,x}$ are ordered lexicographically.
    \item Any pattern in the symmetric set $\mathcal{S}_{x}$ (see Definition~\ref{def:Sx_set}) does not appear in any codeword $\bold{c}$ in $\mathcal{C}_{m,x}$.
    \item The code $\mathcal{C}_{m,x}$ contains all the codewords satisfying the previous three properties.
    \end{enumerate}
Lexicographic ordering of codewords means that they are
ordered in an ascending manner following the rule $0 < 1$ for any bit, and the bit significance reduces from left to right.
\end{definition}

We define the symbol $z$ as the no writing (no transmission) symbol. A run of $x$ consecutive $z$ symbols is denoted by ${\bf{z}}^x$.
To prevent forbidden patterns from appearing across two consecutive codewords, we simply separate any two consecutive LOCO codewords in a stream by ${\bf{z}}^x$. 

\begin{definition}\label{def:z_bridging_ABC}
Given a LOCO code sequence $\{X_n\}$, where binary LOCO codewords are separated by ${\bf{z}}^x$, we obtain $\{Y_n\}$ via NRZ signaling. 
We define three signals $\{A_n\}$, $\{B_n\}$, and $\{C_n\}$ by:

\begin{itemize}
    \item If $X_i = 0$, it is translated to $Y_i=-1$, $A_i=0$, $B_i=1$, $C_i=1$.
    \item If $X_i = 1$, it is translated to $Y_i=1$, $A_i=1$, $B_i=0$, $C_i=1$.
    \item If $X_i = z$, it is translated to $Y_i=0$, $A_i=0$, $B_i=0$, $C_i=0$.
\end{itemize}
\end{definition}

\begin{theorem}\label{thm:2}
The power spectral density of the modulation sequence $\{Y_n\}$ of a LOCO code $\mathcal{C}_{m,x}$ with ${\bf{z}}^x$ bridging pattern and NRZ signaling is given by
\begin{align}\label{eq:SY_AC}
    S_Y(D) = 4S_A(D)-S_C(D).
\end{align}
\end{theorem}
\begin{IEEEproof}
It follows from Definition~\ref{def:z_bridging_ABC} that
\begin{align}
    S_C(D) =& \sum_{j=-\infty}^\infty \overline{\mathbb{E}}[C_\ell C_{\ell+j}]\cdot D^j = \sum_{j=-\infty}^\infty \overline{\mathbb{P}}[C_\ell=C_{\ell+j}=1]\cdot D^j \nonumber\\
    =& \sum_{j=-\infty}^\infty ( \overline{\mathbb{P}}[Y_\ell=Y_{\ell+j}=1] + \overline{\mathbb{P}}[Y_\ell=Y_{\ell+j}=-1] \nonumber \\ 
    &\hspace{+1.5em} + \overline{\mathbb{P}}[Y_\ell = 1, Y_{\ell+j}=-1] + \overline{\mathbb{P}}[Y_\ell = -1, Y_{\ell+j}= 1])\cdot D^j.
\end{align}
Similarly,
\begin{align}
    S_A(D)=\sum_{j=-\infty}^\infty \overline{\mathbb{P}}[Y_\ell=Y_{\ell+j}=1]\cdot D^j
    \text{, and }
    S_B(D)= \sum_{j=-\infty}^\infty \overline{\mathbb{P}}[Y_\ell=Y_{\ell+j}=-1]\cdot D^j.
\end{align}
Consequently, 
\vspace{-0.5em}\begin{align}\label{eq:SY_ABC}
    S_Y(D) =& \sum_{j=-\infty}^\infty \overline{\mathbb{E}}[Y_\ell Y_{\ell+j}]\cdot D^j \nonumber \\ 
    =& \sum_{j=-\infty}^\infty ( \overline{\mathbb{P}}[Y_\ell=Y_{\ell+j}=1] + \overline{\mathbb{P}}[Y_\ell=Y_{\ell+j}=-1] \nonumber \\
    &\hspace{+1.5em} - \overline{\mathbb{P}}[Y_\ell=1, Y_{\ell+j}=-1] - \overline{\mathbb{P}}[Y_\ell=-1, Y_{\ell+j}=1])\cdot D^j \nonumber\\
    =&2[S_A(D)+S_B(D)]-S_C(D).
\end{align}

Because of the inherent symmetry of LOCO codes, we have $S_A(D) = S_B(D)$. Therefore,~(\ref{eq:SY_ABC}) can be simplified as~(\ref{eq:SY_AC}).
\end{IEEEproof}

Observe that the OSTM ${\bf{G}}_C(D)$ of $\{C_n\}$ is simply given by
\begin{align}
    {\bf{G}}_C(D)=\begin{bmatrix}
    \bold{0} & D\cdot {\bf{I}}_{m-1}\\
    D^{x+1} & \bold{0}
    \end{bmatrix}.
\end{align}

\vspace{+0.05em}
\begin{example}
We illustrate our method for obtaining the PSD of LOCO codes in this example. Consider a code whose set of forbidden patterns is $\{010,101\}$ and all codewords are of length $4$, i.e., $x = 1$ and $m = 4$. We adopt the bridging pattern described for signal $A_n$ in Definition \ref{def:z_bridging_ABC}. In practice, we flip all bits of $A_n$ to eliminate infinite sums in the general solution for the OSTM of $S_A(D)$. Because LOCO codes are symmetric and level-based signaling is being used, the continuous spectrum resulting from this OSTM will still be $S_A(D)$.

In accordance with our method, the initial FSTD is given in Fig.~\ref{fig:LOCO_Init}. Then, if unused states are deleted, only states resulting from the most recent bit being a $1$ are labeled in gray, and states are combined when desirable, the final FSTD becomes Fig.~\ref{fig:LOCO_Fin}.

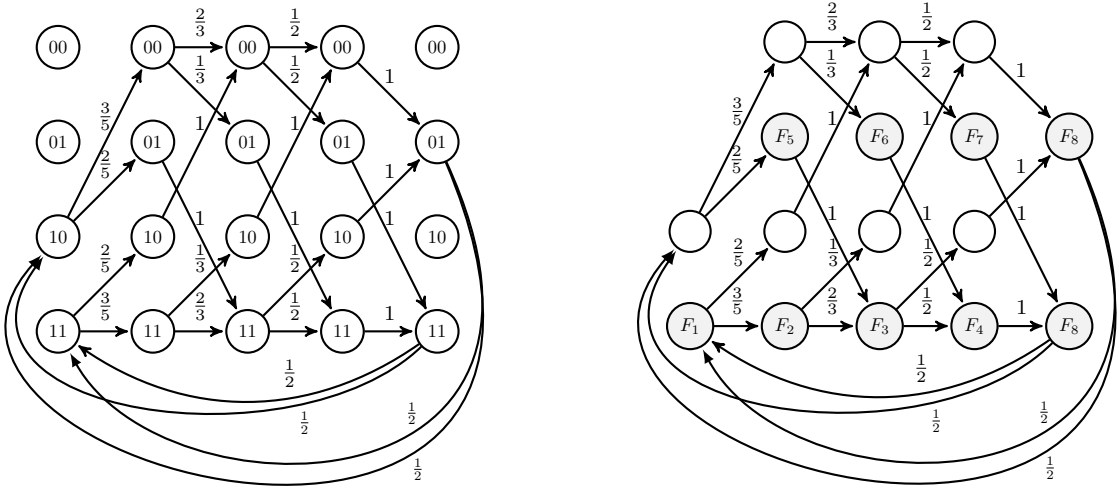
\begin{figure}
\vspace{-0.5em}
\hspace{-2.7em}\begin{subfigure}[b]{.5\textwidth}
    \centering
    \begin{tikzpicture}[->,>=stealth',shorten >=1pt,auto,node distance=1.8cm, thick, scale=0.7, every node/.append style={transform shape}, main node/.style={circle,draw,font=\sffamily\small\bfseries},every edge/.append style={nodes={font=\sffamily\scriptsize}},every path/.append style={nodes={font=\sffamily\scriptsize}}]
    \node[main node] (1) {$00$};
    \node[main node] (2) [below of=1] {$01$};
    \node[main node] (3) [below of=2]{$10$};
    \node[main node] (4) [below of=3]{$11$};
    \node[main node] (5) [right of=1] {$00$};
    \node[main node] (6) [below of=5] {$01$};
    \node[main node] (7) [below of=6]{$10$};
    \node[main node] (8) [below of=7]{$11$};
    \node[main node] (9) [right of=5] {$00$};
    \node[main node] (10) [below of=9] {$01$};
    \node[main node] (11) [below of=10]{$10$};
    \node[main node] (12) [below of=11]{$11$};
    \node[main node] (13) [right of=9] {$00$};
    \node[main node] (14) [below of=13]{$01$};
    \node[main node] (15) [below of=14]{$10$};
    \node[main node] (16) [below of=15]{$11$};
    \node[main node] (17) [right of=13] {$00$};
    \node[main node] (18) [below of=17] {$01$};
    \node[main node] (19) [below of=18]{$10$};
    \node[main node] (20) [below of=19]{$11$};
    \path[every node/.style={font=\sffamily\small}]
        (3) edge node [above] {$\frac{3}{5}$} (5)
            edge node[above] {$\frac{2}{5}$} (6)
        (4) edge node [above] {$\frac{2}{5}$} (7)
            edge node[above] {$\frac{3}{5}$} (8)
        (5) edge node [above] {$\frac{2}{3}$} (9)
            edge node[above] {$\frac{1}{3}$} (10)
        (6) edge node [above] {$1$} (12)
        (7) edge node [above] {$1$} (9)
        (8) edge node [above] {$\frac{1}{3}$} (11)
            edge node[above] {$\frac{2}{3}$} (12)
        (9) edge node [above] {$\frac{1}{2}$} (13)
            edge node[above] {$\frac{1}{2}$} (14)
        (10) edge node[above] {$1$} (16) 
        (11) edge node [above] {$1$} (13)
        (12) edge node [above] {$\frac{1}{2}$} (15)
            edge node[above] {$\frac{1}{2}$} (16) 
        (13) edge node [above] {$1$} (18)
        (14) edge node[above] {$1$} (20) 
        (15) edge node [above] {$1$} (18)
        (16) edge node[above] {$1$} (20) 
        (20) edge [bend left=35] node [above,pos=0.38] {$\frac{1}{2}$} (4);
    \path [line] (20) .. controls (4.5,-8) and (-2.5,-7) .. (3) node[transition,pos=0.23,below] {$\frac{1}{2}$};
    \path [line] (18) .. controls (11.5,-12) and (-4,-8) .. (3) node[transition,pos=0.3,below] {$\frac{1}{2}$};
    \path [line] (18) .. controls (10.5,-9) and (2,-9) .. (4) node[transition,pos=0.4,above] {$\frac{1}{2}$};
    \end{tikzpicture}
    \vspace{-6.5em}
    \caption{Initial $\mathcal{C}_{4,1}$ FSTD. (Steps 1--3)}\label{fig:LOCO_Init}
\end{subfigure}
\begin{subfigure}[b]{.5\textwidth}
    \centering
    \begin{tikzpicture}[->,>=stealth',shorten >=1pt,auto,node distance=1.8cm, thick, scale=0.7, every node/.append style={transform shape}, main node/.style={circle,draw,font=\sffamily\small\bfseries},every edge/.append style={nodes={font=\sffamily\scriptsize}},every path/.append style={nodes={font=\sffamily\scriptsize}}]
    \node[main node] (3) {\textcolor{white}{$D$}};
    \node[main node] (4) [below of=3,fill=gray!10]{\small$F_1$};
    \node[main node] (7) [right of=3]{\textcolor{white}{$D$}};
    \node[main node] (6) [above of=7,fill=gray!10]{\small$F_5$};
    \node[main node] (5) [above of=6] {\textcolor{white}{$D$}};
    \node[main node] (8) [below of=7,fill=gray!10]{\small$F_2$};
    \node[main node] (9) [right of=5] {\textcolor{white}{$D$}};
    \node[main node] (10) [below of=9,fill=gray!10]{\small$F_6$};
    \node[main node] (11) [below of=10]{\textcolor{white}{$D$}};
    \node[main node] (12) [below of=11,fill=gray!10]{\small$F_3$};
    \node[main node] (13) [right of=9] {\textcolor{white}{$D$}};
    \node[main node] (14) [below of=13,fill=gray!10]{\small$F_7$};
    \node[main node] (15) [below of=14]{\textcolor{white}{$D$}};
    \node[main node] (16) [below of=15,fill=gray!10]{\small$F_4$};
    \node[main node] (18) [right of=14,fill=gray!10]{\small$F_8$};
    \node[main node] (20) [right of=16,fill=gray!10]{\small$F_8$};
    \path[every node/.style={font=\sffamily\small}]
        (3) edge node [above] {$\frac{3}{5}$} (5)
            edge node[above] {$\frac{2}{5}$} (6)
        (4) edge node [above] {$\frac{2}{5}$} (7)
            edge node[above] {$\frac{3}{5}$} (8)
        (5) edge node [above] {$\frac{2}{3}$} (9)
            edge node[above] {$\frac{1}{3}$} (10)
        (6) edge node [above] {$1$} (12)
        (7) edge node [above] {$1$} (9)
        (8) edge node [above] {$\frac{1}{3}$} (11)
            edge node[above] {$\frac{2}{3}$} (12)
        (9) edge node [above] {$\frac{1}{2}$} (13)
            edge node[above] {$\frac{1}{2}$} (14)
        (10) edge node[above] {$1$} (16) 
        (11) edge node [above] {$1$} (13)
        (12) edge node [above] {$\frac{1}{2}$} (15)
            edge node[above] {$\frac{1}{2}$} (16) 
        (13) edge node [above] {$1$} (18)
        (14) edge node[above] {$1$} (20) 
        (15) edge node [above] {$1$} (18)
        (16) edge node[above] {$1$} (20) 
        (20) edge [bend left=35] node [above,pos=0.38] {$\frac{1}{2}$} (4);
    \path [line] (20) .. controls (4.5,-4.5) and (-2.5,-3.5) .. (3) node[transition,pos=0.23,below] {$\frac{1}{2}$};
    \path [line] (18) .. controls (11.5,-8.5) and (-4,-4.5) .. (3) node[transition,pos=0.3,below] {$\frac{1}{2}$};
    \path [line] (18) .. controls (10.5,-5.5) and (2,-5.5) .. (4) node[transition,pos=0.4,above] {$\frac{1}{2}$};
    \end{tikzpicture}
    \vspace{-6.5em}
    \caption{Final $\mathcal{C}_{4,1}$ FSTD. (Steps 1--6)}\label{fig:LOCO_Fin}
\end{subfigure}
\caption{FSTD used for OSTM generation of the $A$ sub-signal of a $\mathcal{C}_{4,1}$ sequence.}
\vspace{-1.0em}
\end{figure}

The resulting OSTM is then 
\begin{align}
    {\bf{G}}_A(D) = \begin{bmatrix}
    0 & \frac{3}{5}D & 0 & 0 & 0 & 0 & \frac{1}{5}D^3 & \frac{1}{5}D^4\\
    0 & 0 & \frac{2}{3}D & 0 & 0 & 0 & 0 & \frac{1}{3}D^3\\
    0 & 0 & 0 & \frac{1}{2}D & 0 & 0 & 0 & \frac{1}{2}D^2\\
    0 & 0 & 0 & 0 & 0 & 0 & 0 & D\\
    0 & 0 & D & 0 & 0 & 0 & 0 & 0\\
    0 & 0 & 0 & D & 0 & 0 & 0 & 0\\
    0 & 0 & 0 & 0 & 0 & 0 & 0 & D\\
    \frac{1}{2}D & 0 & 0 & 0 & \frac{1}{5}D^2 & \frac{1}{10}D^3 & \frac{1}{10}D^4 & \frac{1}{10}D^5\\
    \end{bmatrix}.
\end{align}

Once the OSTM is found, we use Theorem~\ref{thm:1} to calculate the sub-signal PSD $S_A(D)$, and we then use Theorem~\ref{thm:2} for $S_Y(D)$ and (\ref{eq:Sw(f)}) to calculate the continuous component of the PSD of $\mathcal{C}_{4,1}$ theoretically, see Fig.~\ref{fig:cont_LOCO}. A LOCO code does not have a discrete PSD component, and this will be discussed shortly.

\begin{figure}
\centering
\includegraphics[width=10cm]{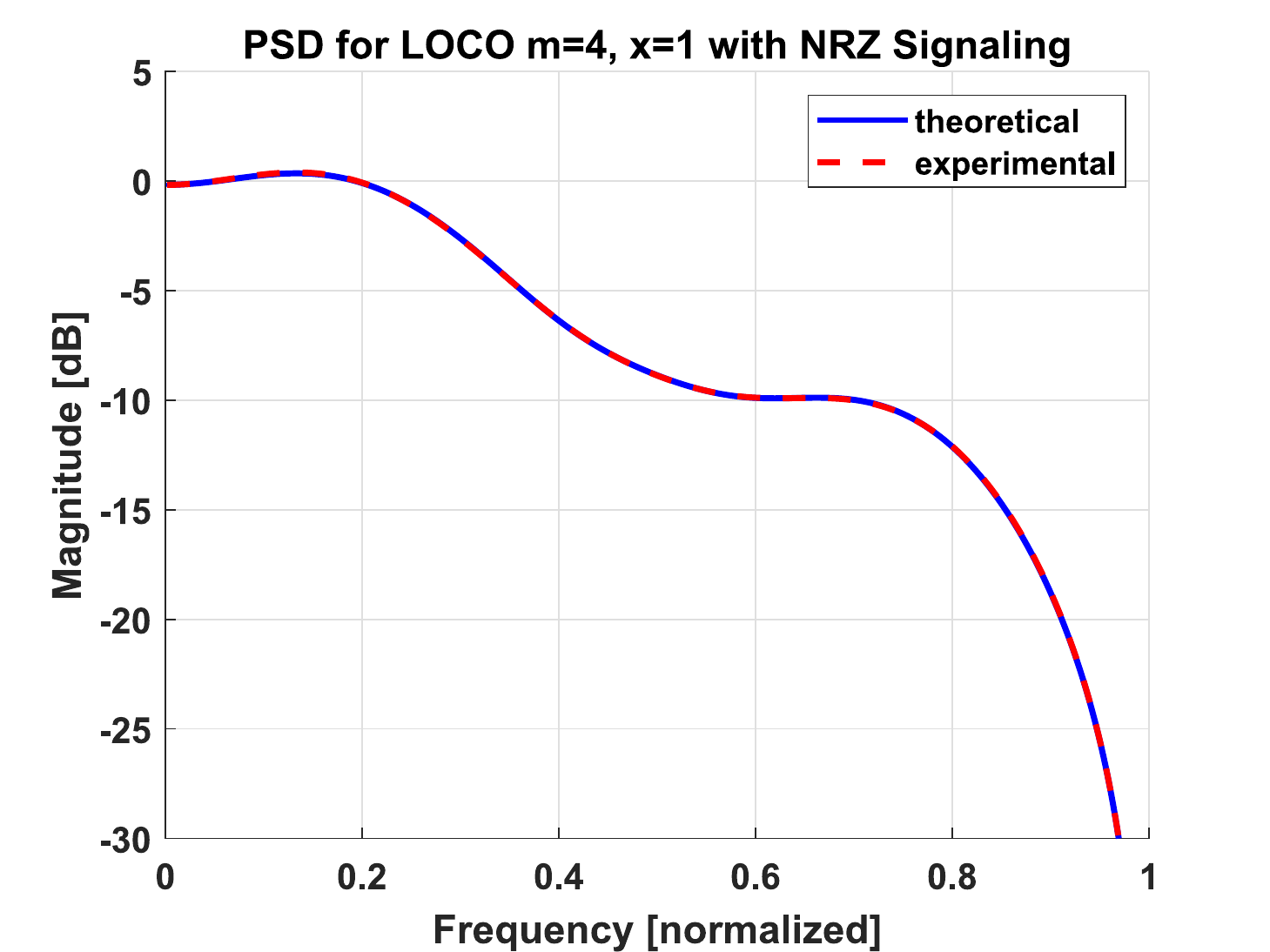}
\caption{The theoretical (OSTM method) versus experimental (Monte-Carlo) continuous component of the PSD for LOCO code $\mathcal{C}_{4,1}$ with NRZ signaling.}\label{fig:cont_LOCO}
\end{figure}

\end{example}

\begin{result}
The general OSTM solution for the $A$ sub-signal of any LOCO code $\mathcal{C}_{m,x}$ defined by the parameters $m$ and $x$, and using ${\bf{z}}^x$ bridging is the $m+xm+x-x^2$ square matrix ${\bf{G}}_A(D)$, which is sparse. The non-zero entries in ${\bf{G}}_A(D)$ can be found using the following rules:

%\begin{align}
%    G(D) = \begin{bmatrix}
%    \begin{matrix}
%        0 & A & 0 & \cdots & 0\\ \vdots & 
%        \begin{matrix}
%            0 & \cdots & \cdots
%        \end{matrix}
%        & \cdots & \cdots& 0\\
%    \end{matrix}
%    & C & 
%    \begin{matrix}
%        0\\ \vdots\\
%    \end{matrix}\\
%    \begin{matrix}
%        0 & 
%        \begin{matrix}
%            \cdots & 0 & B 
%        \end{matrix}
%        & 0 & \cdots & \cdots 
%    \end{matrix}
%    & \cdots & 0\\
%    \begin{matrix}
%        0 & 
%        \begin{matrix}
%            \cdots & \cdots & \cdots 
%        \end{matrix}
%        & . & \cdots & 0 
%    \end{matrix}
%    & E\\
%    \begin{matrix}
%        F & 
%        \begin{matrix}
%            0 & \cdots & 0 
%        \end{matrix}
%        & G
%    \end{matrix}
%    & 0\\
%    \end{bmatrix}
%\end{align}

\begin{itemize}
  \item If $i=m+xm+m-x^2$ and $j=1$, then ${\bf{G}}_A(D)_{i,j} = \frac{1}{2}D$.
  \item If $i=j-1$ and $i<m$, then ${\bf{G}}_A(D)_{i,j} = \lambda_{m-i+1}D$.
  \item If $xm+x-x^2-1<j<m+xm+x-x^2$ and $i<m+1$, then consider the sub-matrix $\bold{A}_{m \times m}$ with non-zero entries ${\bf{A}}_{k,l}$, $1 \leq k,l \leq m$, set as follows.
    \begin{itemize}
        \item If $l=m$ or $l-k>x$, ${\bf{A}}_{k,l}=(1-\lambda_{m-k+1})(1-\lambda_{m-l+1})D^{l-k+1} \underbrace{\left( \prod_{g=m-l+2}^{m-x-1} \lambda_{g} \right)}_{\text{if } l>x+2}$.
    \end{itemize}
  \item If $m<i<m+xm-x-x^2+1$ and $x+1<j<xm-x^2+2$, then ${\bf{G}}_A(D)_{i,j} = D$.
  \item If $m+xm-x-x^2<i<m+xm+x-x^2$ and $m+xm-x-x^2+1<j<m+xm+x-x^2+1$, then ${\bf{G}}_A(D)_{i,j} = D$.
  \item If $i=m+xm+m-x^2$ and $xm+x-x^2-1<j<m+xm+x-x^2$, then consider the sub-vector $\bold{a}_{1 \times m}$ with entries ${\bf{a}}_{l}$, $1 \leq l \leq m$, set as follows.
    \begin{itemize}
        \item ${\bf{a}}_{l}=\frac{1}{2} (1-\lambda_{m-l+1}) \prod_{g=m-l+2}^{m} \lambda_g$.
    \end{itemize}
\end{itemize}
In those rules,
\vspace{-0.1em}\begin{align}
    \lambda_{a}=\frac{N_1(a,x)}{\frac{1}{2}N(a,x)}.
\end{align} 
The notation $N(m,x)$ and $N_1(m,x)$ refers to the cardinality of $\mathcal{C}_{m,x}$ and the cardinality of Group 1 in $\mathcal{C}_{m,x}$ (the number of codewords starting with $00$ from the left), respectively, adopted from \cite{ahh_loco}.

\end{result}

\begin{figure}
    \begin{subfigure}[b]{.5\textwidth}
        \centering
        \includegraphics[width=3.25in]{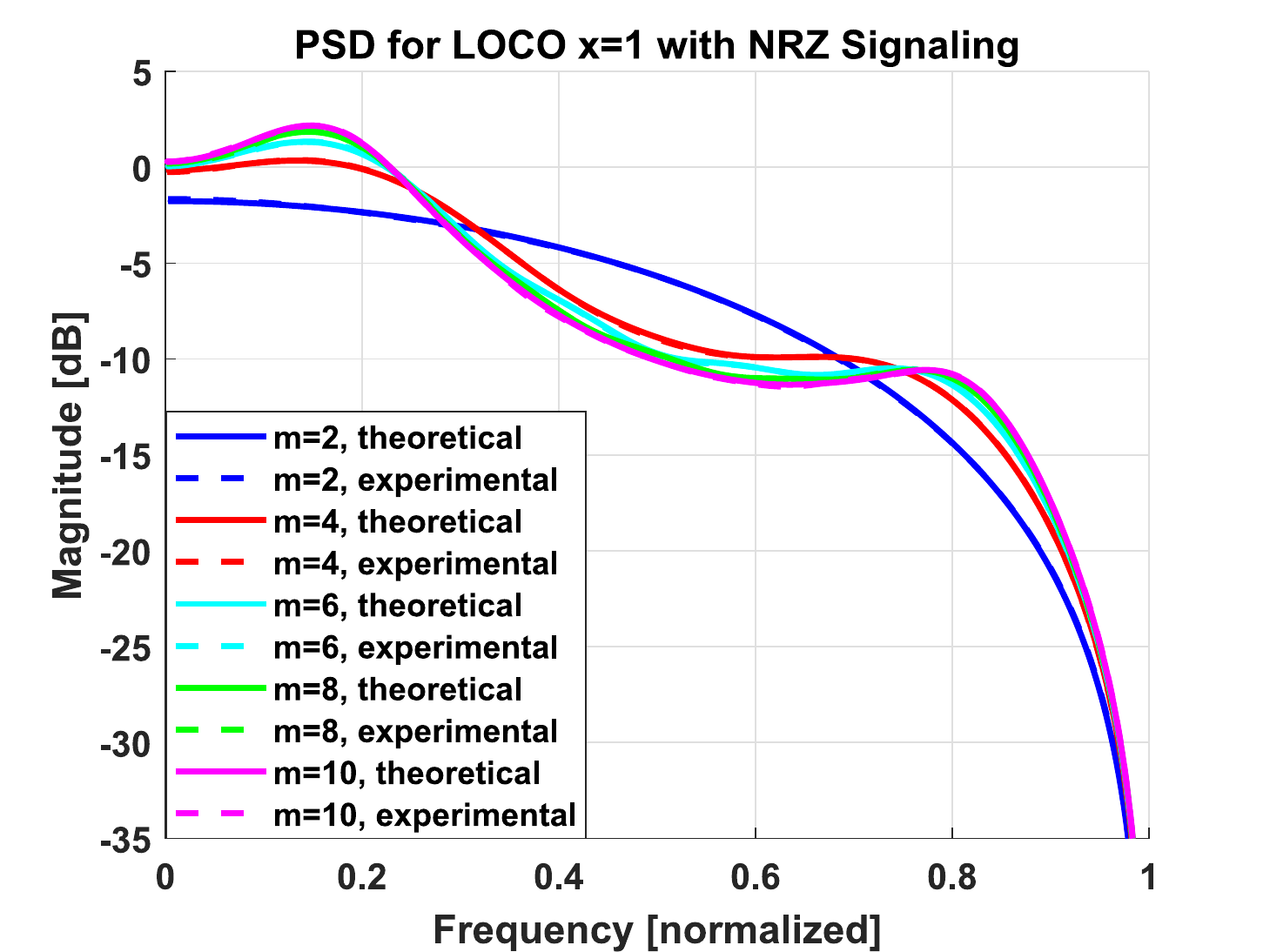}
    \end{subfigure}
    \begin{subfigure}[b]{.5\textwidth}
        \centering
        \includegraphics[width=3.25in]{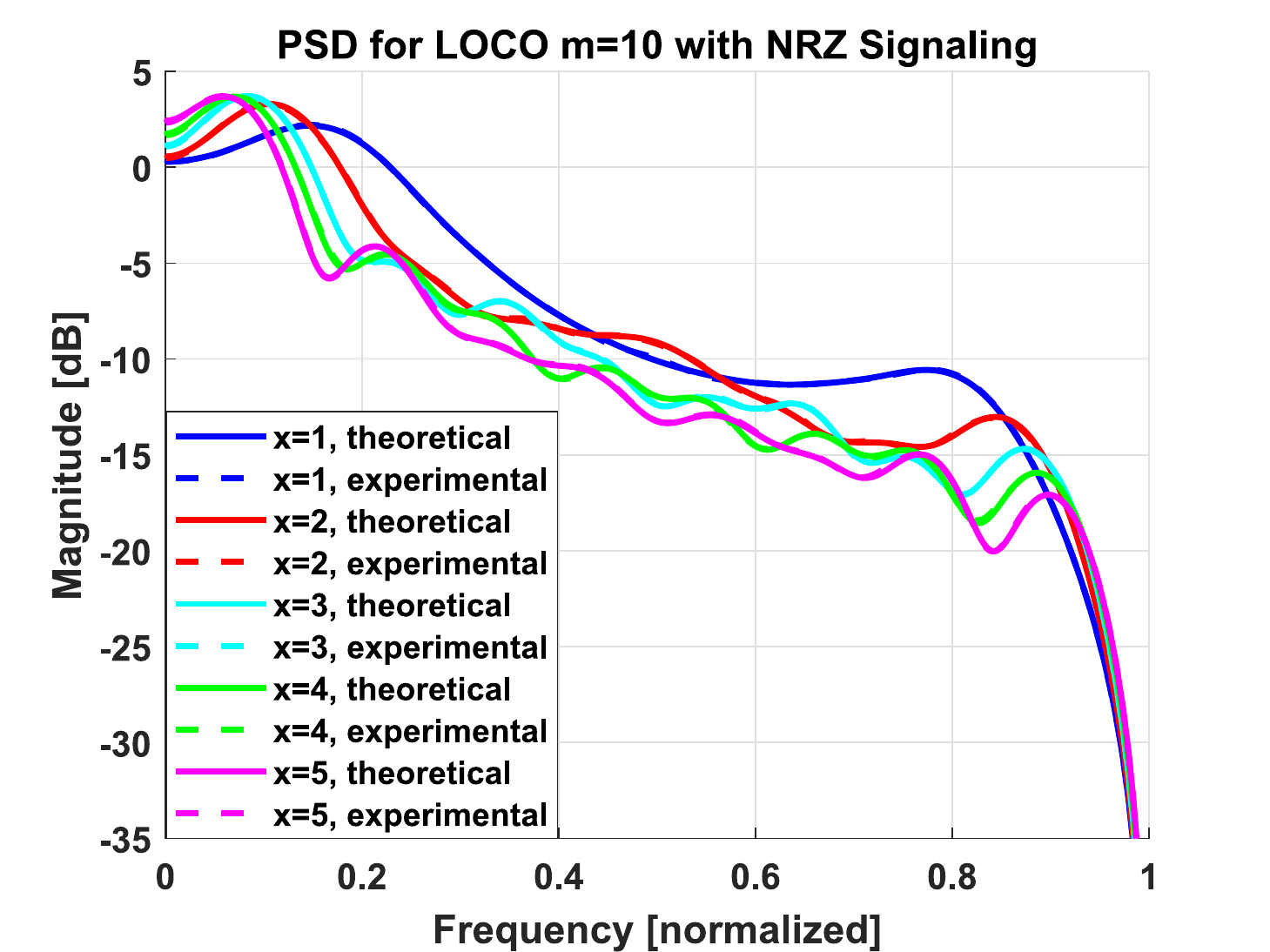}
    \end{subfigure}
    \caption{Comparisons of the continuous component of various $\mathcal{C}_{m,x}$ spectra.}
    \label{fig:LOCOspectra}
\end{figure}

Since LOCO codes are symmetric, there is no discrete power at frequency $0$ (DC) when NRZ signaling is adopted since $0$'s ($-1$'s) and $1$'s ($+1$'s) are equally likely to occur, i.e., $\mathbb{E}[Y_n] = 0$. Symmetric coding with NRZ signaling leads to the case where the auto-correlation function is equal to the auto-covariance function. Consequently, there is no discrete component at all in the PSD for all frequencies. Once again, the continuous PSD at frequency $0$ increases and the number of spectral peaks increases for the same $m$ as $x$ increases. These spectral characteristics along with others are demonstrated in Fig.~\ref{fig:LOCOspectra}.

\begin{table}
\vspace{-0.2em}
\centering
 \caption{$3$dB Bandwidth for Various LOCO Codes}\label{tb:loco}
 \begin{tabular}{||p{3.0cm}||c|c|c|c|c|c|c|c|c||} 
 \hline
 \multicolumn{10}{||c||}{LOCO $3$dB Bandwidth Measurements} \\
 \hline\hline
 $x$ [bits] & $1$ & $1$ & $1$ & $1$ & $1$ & $2$ & $3$ & $4$ & $5$  \\ 
 \hline
 $m$ [bits] & $2$ & $4$ & $6$ & $8$ & $10$ & $10$ & $10$ & $10$ & $10$ \\
 \hline
 $3$dB BW [normalized] & $0.868$ & $0.644$ & $0.582$ & $0.568$ & $0.558$ & $0.412$ & $0.327$ & $0.283$ & $0.246$ \\
 \hline
 \end{tabular}
 \vspace{-0.3em}
 \end{table}
 
For the examples in Fig.~\ref{fig:LOCOspectra}, the $3$dB bandwidths are presented in Table~\ref{tb:loco}. Here, the $3$dB bandwidth decreases for the same $m$ as $x$ increases since the minimum and average time separations between consecutive transitions in $W(t)$ become bigger.

% %%%%%%%%%%%%%%%%%%%%%%%%%%%%%%%%%%%%%
% \section{PSD for Clocked Constrained Codes}\label{sec_clock}

% \nt{I suggest moving this section to appendix. The descriptions on the algorithm and how it fit into C-LOCO and CA-LOCO codes are heavy. I don't think it worth. Appendix is better, by giving a general idea and a pseudo-code.}
% \jc{Works for me.}

%%%%%%%%%%%%%%%%%%%%%%%%%%%%%%%%%%%%%
\section{Conclusion}\label{sec_conc}

We introduced a method for deriving the power spectra of constrained codes associated with level-based signaling. We applied this method to infinite-length $\mathcal{A}_x$-constrained and $\mathcal{S}_x$-constrained codes. We then generalized this method to address finite-length challenges in deriving power spectra, particularly how to find the OSTM, how to address cyclostationarity, and how to handle the periodic component of the auto-correlation function. We applied the generalized method to finite-length A-LOCO and LOCO codes. We discussed important spectral properties of these classes of codes based on our derivations. Our theoretical PSD results perfectly match the experimental PSD results for all codes. We present another method for deriving the spectra along with an algorithmic approach for self-clocked codes in the appendices. We suggest that our methods and ideas can provide helpful insights to data storage and data transmission engineers regarding different classes of constrained codes they plan to employ.

%%%%%%%%%%%%%%%%%%%%%%%%%%%%%%%%%%%%%
\section*{Acknowledgment}\label{sec_ack}

This research was supported in part by NSF under Grant CCF 1717602 and in part by AFOSR under Grant FA 9550-17-1-0291.

%%%%%%%%%%%%%%%%%%%%%%%%%%%%%%%%%%%%%
\appendices

\section{Alternate Method for Obtaining the PSD of $\mathcal{A}_x$- and $\mathcal{S}_x$-Constrained Codes}\label{ap:A}

Instead of altering the NRZI-focused method seen in \cite{gall_psd} for a code with NRZ signaling as we did in this paper, the relationship between NRZI signals and their base binary sequence can be used to modify the constraints of the base sequence.
NRZI signaling is transition-based, with each transition (from $-1$ to $+1$ or $+1$ to $-1$) being caused by a $1$ in the base sequence. 
Therefore, in order to use the unaltered method, all we need to do is to generate a constrained sequence whose NRZI signal is equivalent to the NRZ signal of the desired sequence.

We focus here on infinite-length codes. We start from the desired binary sequence, and perform ``\textit{inverse NRZI}'' operation on it. In particular, we take the bit-wise difference between each two adjacent bits in the desired sequence, resulting in a new binary sequence whose NRZI signal is the NRZ signal we are after, assuming the initial signal level is the same. What remains is that we identify the new constraint in this bit-wise difference sequence in order to build its FSTD and OSTM, and thus derive its PSD. This is done as follows for the codes we discuss here. Suppose the initial level is $-1$.
\begin{itemize}
\item For $\mathcal{A}_x$-constrained codes, the new constraint is that separation between $1$'s indexed by $2j$ and $2j+1$ (resp., $2j+1$ and $2j+2$) is unrestricted (resp., at least $x$ $0$'s), where $j \in \mathbb{N}$.

\item For $\mathcal{S}_x$-constrained codes, the new constraint is that separation between $1$'s indexed by $j$ and $j+1$ is at least $x$ $0$'s, where $j \in \mathbb{N}$. This is exactly the $d$ constraint in an RLL code for $d=x$.
\end{itemize}

The unaltered method in \cite{gall_psd} is then used to derive the PSD via the OSTM built using the new constraint. Observe that we are discussing here the continuous component of the PSD. The delta function at DC in the case of asymmetric codes is computed separately as discussed in Section~\ref{sec_finite}. Observe that for an asymmetric sequence, the initial level is quite important for generating the desired NRZ signal. An example of the FTSDs for an $\mathcal{A}_1$-constrained sequence and its bit-wise difference sequence is provided in Fig.~\ref{fig:FSTD_bwdiff}.

%For an asymmetric sequence, this unknown can result in an inaccuracy of calculating the power level at DC ($f=0$). Since asymmetry results in a discrete component appearing at DC, the power level is already required to be calculated separately from this theoretical method of estimating the PSD, so no further consideration is taken for this truth.

\tikzset{node distance=1.4cm,
every state/.style={semithick,minimum size=1pt},
initial text={},
double distance=2pt,
every edge/.style={ draw,
->,>=stealth', auto, semithick}}
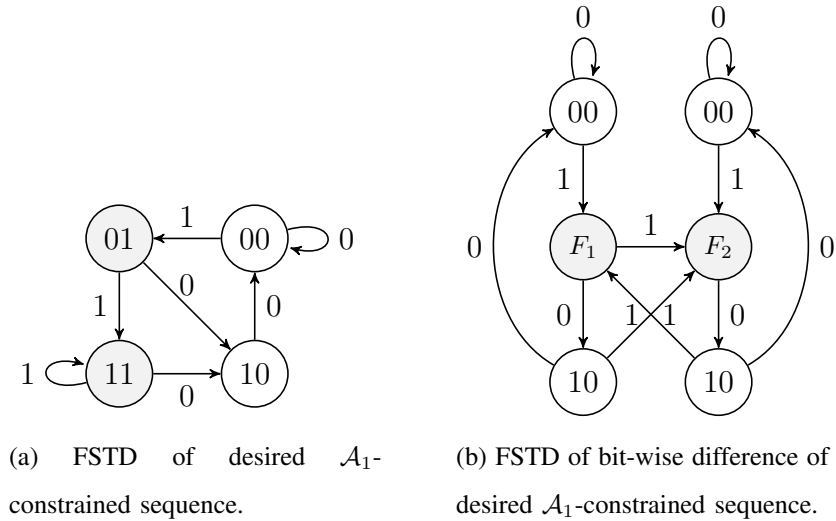
\begin{figure}
\centering
\begin{subfigure}[t]{.3\textwidth}
\begin{tikzpicture}[node distance=1.8cm]
\node[state] (A) {$00$};
\node[state, fill=gray!10, left of=A] (B) {$01$}; 
\node[state, fill=gray!10, below of=B] (C) {$11$};
\node[state, right of=C] (D) {$10$};
\draw (A) edge[loop right] node {$0$} (A);
\draw (A) edge node[above] {$1$} (B);
\draw (B) edge node[left] {$1$} (C);
\draw (B) edge node[above] {$0$} (D);
\draw (C) edge[loop left] node {$1$} (C);
\draw (C) edge node[below] {$0$} (D);
% \draw (D) edge[bend right=30] node[right] {0} (A);
\draw (D) edge node[right] {$0$} (A);
\end{tikzpicture}
\caption{FSTD of desired $\mathcal{A}_1$-constrained sequence.}
\end{subfigure}
\hspace{+2.0em}\begin{subfigure}[t]{.3\textwidth}
\begin{tikzpicture}[node distance=1.8cm]
\node[state] (A) {$00$};
\node[state, right of=A] (B) {$00$};
\node[state, fill=gray!10, below of=A] (C) {\small$F_1$};
\node[state, fill=gray!10, below of=B] (D) {\small$F_2$};
\node[state, below of=C] (E) {$10$};
\node[state, below of=D] (F) {$10$};
\draw (A) edge[loop above] node {$0$} (A);
\draw (A) edge node[left] {$1$} (C);
\draw (B) edge[loop above] node {$0$} (B);
\draw (B) edge node[right] {$1$} (D);
\draw (C) edge node[above] {$1$} (D);
\draw (C) edge node[left] {$0$} (E);
\draw (D) edge node[right] {$0$} (F);
\draw (E) edge node[left] {$1$} (D);
\draw (E) edge[bend left=60] node[left] {$0$} (A);
\draw (F) edge node[right] {$1$} (C);
\draw (F) edge[bend right=60] node[right] {$0$} (B);
\end{tikzpicture}
\caption{FSTD of bit-wise difference of desired $\mathcal{A}_1$-constrained sequence.}
\end{subfigure}
\caption{FSTDs considered for using the unaltered method \cite{gall_psd} for an $\mathcal{A}_1$-constrained sequence.}
\label{fig:FSTD_bwdiff}
\vspace{-1.0em}
\end{figure}

\begin{result}
By exploring FSTDs and OSTDs of the bit-wise difference of various $\mathcal{A}_x$-constrained sequences, it can be found that the alternative OSTM for any $\mathcal{A}_x$-constrained sequence is
\begin{align}
    {\bf{G}}(D) = 
    \begin{bmatrix}
    0 & \sum_{k=1}^{\infty}(\frac{1}{2})^{k}D^{k} \\
    \sum_{k=1}^{\infty}(\frac{1}{2})^{k}D^{x+k} & 0 
    \end{bmatrix} = 
    \begin{bmatrix}
    0 & \frac{D}{2-D} \\
    \frac{D^{1+x}}{2-D} & 0 
    \end{bmatrix}.
\end{align}

Similarly, the alternative OSTM for any $\mathcal{S}_x$-constrained sequence is
\begin{align}
    {\bf{G}}(D) = 
    \begin{bmatrix}
    \sum_{k=1}^{\infty}(\frac{1}{2})^{k}D^{x+k}
    \end{bmatrix} =
    \begin{bmatrix}
    \frac{D^{1+x}}{2-D}
    \end{bmatrix}.
\end{align}
\end{result}

\section{Generating the OSTM of Clocked Constrained Codes}\label{ap:B}

\begin{definition}~\cite{ahh_loco}~\cite{ahh_aloco}
A self-clocked LOCO (C-LOCO) code $\mathcal{C}_{m,x}^{\textup{clk}}$ is defined by
\begin{align}
    \mathcal{C}_{m,x}^{\textup{clk}} = \mathcal{C}_{m,x} \setminus \{{\bf 0}^m, {\bf 1}^m\}, \text{ where }m\geq 2.
\end{align}
Furthermore, a self-clocked A-LOCO (CA-LOCO) code $\mathcal{AC}_{m,x}^{\textup{clk}}$ is defined by
\begin{align}
    \mathcal{AC}_{m,x}^{\textup{clk}} = \mathcal{AC}_{m,x} \setminus \{{\bf 0}^m, {\bf 1}^m\}, \text{ where } m\geq 2.
\end{align}
\end{definition}

Self-clocked codes impose another constraint on the stream generation, causing the method used in Section~\ref{sec_finiteFSTD} for finding the FSTD and OSTM of the code to require more than the maximum $(m+x)2^{x+1}$ states introduced. One idea for handling this additional complexity is to utilize the breadth first search (BFS) algorithm to properly construct the code OSTD and OSTM. Details on the BFS algorithm are not discussed in this paper for brevity.

\begin{algorithm}
\caption{Deriving OSTD of Self-Clocked Codes}
\begin{algorithmic}
\State \textbf{Inputs:} ${\bf{P}}$, ${\bf{V}}$, ${\bf{\text{target}}}$, and $k_{\text{eff}}^{\textup{clk}}$.
\State \textbf{for} $d' = 1:(k_{\text{eff}}^{\textup{clk}}+1)$ \textbf{do}
\State \hspace{2ex} ${\bf{V}}_1 = {\bf{P}}\times {\bf{V}}$. \textit{(matrix multiplication)}
\State \hspace{2ex} ${\bf{V}}_2 = {\bf{\text{target}}} \hspace{+0.2em}.\hspace{-0.2em}\times {\bf{V}}_1$. \textit{(element-wise multiplication)}
\State \hspace{2ex} Record $d'$ and all the non-zero entries in ${\bf{V}}_2$.
\State \hspace{2ex} ${\bf{V}} = {\bf{V}}_1 \hspace{+0.2em}.\hspace{-0.2em}\times(\sim\hspace{-0.2em}{\bf{V}}_2)$. \textit{($\sim\hspace{-0.2em}{\bf{V}}_2$ is the bit flipping matrix of the binary version of ${\bf{V}}_2$)}
\State \textbf{end for}
\State Recover the recorded numbers of steps (values of $d'$ for ${\bf{V}}_2$) as labels on one graph.
\State Recover the recorded probabilities (non-zero entries in ${\bf{V}}_2$) as labels on another graph.
\State \textbf{Output:} OSTD of the code with number of steps and corresponding probability for each edge.
\end{algorithmic}
\label{alg_bsf}
\end{algorithm}

To make use of the BFS algorithm, first note that for every codeword in $\mathcal{C}_{m,x}^{\textup{clk}}$ or $\mathcal{AC}_{m,x}^{\textup{clk}}$, there exists at least one transition for self-clocking. Thus, we can define the maximum number of successive symbols between two consecutive $1$'s as $k_{\text{eff}}^{\textup{clk}} = 2(m-1)+x$, see \cite{ahh_loco} and \cite{ahh_aloco}. 
We define the following variables as inputs to the BFS algorithm: 
\begin{itemize}
    \item Derive the stochastic probability $t\times t$ matrix ${\bf{P}}$, where $t$ is the number of states in the FSTD and ${\bf{P}}_{ij}$ is the probability of transiting from state $j$ to state $i$ in $1$ step.
    \item Define a column vector ${\bf{w}}$ of length $t$, where ${\bf{w}}_i=1$ if and only if all the incoming edges of vertex $i$ in the FSTD have label $1$, otherwise ${\bf{w}}_i=0$.
    \item Define a $t\times t$ matrix ${\bf{V}}$, where ${\bf{V}}_{ij} = 1$ if and only if $i=j$ and ${\bf{w}}_i=1$, otherwise ${\bf{V}}_{ij} = 0$. Each column of ${\bf{V}}$ has at most one non-zero entry. We start our BFS with ${\bf{V}}$ and update its value after we walk (transit) in the FSTD.
    \item Define a $t\times t$ matrix ${\bf{\text{target}}}={\bf{w}}{\bf{w}}^\mathsf{T}$, i.e., ${\bf{\text{target}}}_{ij} = 1$ if and only if ${\bf{w}}_j = 1$ and ${\bf{w}}_i=1$, otherwise ${\bf{\text{target}}}_{ij} = 0$. In the algorithm, when we hit a non-zero entry in ${\bf{\text{target}}}$, we find a possible $00\dots01$ sequence and its associated probability. 
\end{itemize}

The BFS algorithm is Algorithm~\ref{alg_bsf}.

%%%%%%%%%%%%%%%%%%%%%%%%%%%%%%%%%%%%%

%%%%%%%%%%%%%%%%%%%%%%%%%%%%%%

\end{document}